\documentclass[12pt, a4paper]{article}
\usepackage[english]{babel}
\usepackage[T1]{fontenc}
\usepackage[scaled]{helvet}
\renewcommand*\familydefault{\sfdefault}
\usepackage{eurosym}
\usepackage[dvips]{graphics}
\usepackage{amsmath, amssymb}
\usepackage[usenames,dvipsnames,svgnames,table]{xcolor}
\usepackage{subfig}
\usepackage{array,booktabs}
\usepackage{times}
\usepackage{multicol}
\usepackage{afterpage}
\usepackage{xargs} 
\usepackage{ifthen}
\usepackage{pifont}
\usepackage{setspace}
\usepackage{stmaryrd}
\usepackage{wrapfig,framed}
\usepackage{floatrow} 
\usepackage{caption}%[labelfont={color=ocre,bf}]
\usepackage{mathabx}
\usepackage{enumitem} 
\usepackage{colortbl} 
\usepackage{lipsum}
\usepackage{microtype}                                                          
\usepackage[cm]{fullpage}
\usepackage{ulem} 
\renewcommand{\em}{\it}
\usepackage{url}
\usepackage{hyperref} 
\usepackage{geometry}
\usepackage{fancyhdr}
\usepackage{footnote} 
\usepackage[super,nomove]{cite}

\DeclareColorBox{shaded}{\colorbox{yellow!20}}
\addto\captionsenglish{%
}

\makeatletter
\renewcommand\section{\@startsection{section}{2}{-0.9cm}%
	{0.4cm }%
	{0.3cm }%
	{\reset@font\large\bfseries}}
\makeatother

\newcommand{\figref}[1]{\hyperref[#1]{Fig.\ref{#1}}}

\definecolor{aenlever}{RGB}{100,75,100}
\definecolor{newcol}{RGB}{0,75,100}
\colorlet{cfond}{black!5}
\colorlet{newcolC}{newcol}
\colorlet{newcolL}{ForestGreen}
\colorlet{citecolorSAFE}{ForestGreen}
\colorlet{citecolor}{ForestGreen}
\colorlet{newcolU}{newcol}
\definecolor{crimson}{RGB}{220,20,60}
\colorlet{mycitecolor}{crimson}
\colorlet{mathcolor}{gray}
\definecolor{turquoise4}{RGB}{0,134,139}
\definecolor{cadetblue4}{RGB}{83,134,139}
\definecolor{tan4}{RGB}{139,90,43}
\definecolor{red4}{RGB}{139,0,0} %% presque marron
\definecolor{indianred}{RGB}{176,23,31}
\definecolor{crimson}{RGB}{220,20,60}
\definecolor{goldenrod1}{RGB}{255,193,37}
\definecolor{royalblue4}{RGB}{39,64,139}
\definecolor{azure4}{RGB}{131,139,139}
\definecolor{azure5}{RGB}{105,111,111}
\definecolor{darkolivegreen}{RGB}{85,107,47}
\definecolor{emeraldgreen}{RGB}{0,201,87}

\hypersetup{colorlinks=true, linkcolor=Black!90,urlcolor=OrangeRed,
  citecolor=citecolor}

\newcommand{\citetodo}[1]{{\color{red}\bf [??]}}

\newcommand{\citeal}[1]{{\color{citecolor}\hspace{-2pt}\textsuperscript{[}\cite{#1}\textsuperscript{]}}}

\newcommand{\citemycite}[2]{{\color{citecolor}\textsuperscript{[}\cite{#1}{\color{mycitecolor}\colorlet{citecolor}{mycitecolor}\textsuperscript{\citepunct}
\cite{#2}\colorlet{citecolor}{citecolorSAFE}}\textsuperscript{]}}}

\newcommand{\myciteOL}[1]{{\color{mycitecolor}\colorlet{citecolor}{mycitecolor}[\citenum{#1}]\colorlet{citecolor}{citecolorSAFE}}}

\newcommand{\myurl}[1]{\scalebox{0.8}[1.1]{\textls[-50]{\url{#1}}}}

\usepackage{multibib}
\newcites{latex}{\LaTeX-Literature}%  \citelatex, \nocitelatex, ...
\usepackage{bibentry}

\RequirePackage{tikz} 
\usetikzlibrary{snakes,backgrounds,arrows,matrix,petri,shapes} 
\RequirePackage{extarrows}
\RequirePackage{fourier-orns}

\def\ie{{\it i.e.}~} 
 
\def\eg{{\it e.g.}~}

\newcommand{\Ai}[0]{\scalebox{0.4}{\begin{picture}(0,0)%
\includegraphics{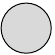}%
\end{picture}%
\setlength{\unitlength}{4144sp}%
\begingroup\makeatletter\ifx\SetFigFont\undefined%
\gdef\SetFigFont#1#2#3#4#5{%
  \reset@font\fontsize{#1}{#2pt}%
  \fontfamily{#3}\fontseries{#4}\fontshape{#5}%
  \selectfont}%
\fi\endgroup%
\begin{picture}(398,398)(2097,-935)
\put(2206,-826){\makebox(0,0)[lb]{\smash{{\SetFigFont{20}{24.0}{\familydefault}{\mddefault}{\updefault}{\color[rgb]{0,0,0}$i$}%
}}}}
\end{picture}%
}}
\newcommand{\Aj}[0]{\scalebox{0.4}{\begin{picture}(0,0)%
\includegraphics{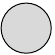}%
\end{picture}%
\setlength{\unitlength}{4144sp}%
\begingroup\makeatletter\ifx\SetFigFont\undefined%
\gdef\SetFigFont#1#2#3#4#5{%
  \reset@font\fontsize{#1}{#2pt}%
  \fontfamily{#3}\fontseries{#4}\fontshape{#5}%
  \selectfont}%
\fi\endgroup%
\begin{picture}(398,398)(2097,-935)
\put(2206,-826){\makebox(0,0)[lb]{\smash{{\SetFigFont{20}{24.0}{\familydefault}{\mddefault}{\updefault}{\color[rgb]{0,0,0}$j$}%
}}}}
\end{picture}%
}}
\newcommand{\Aone}[0]{\scalebox{0.4}{\begin{picture}(0,0)%
\includegraphics{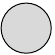}%
\end{picture}%
\setlength{\unitlength}{4144sp}%
\begingroup\makeatletter\ifx\SetFigFont\undefined%
\gdef\SetFigFont#1#2#3#4#5{%
  \reset@font\fontsize{#1}{#2pt}%
  \fontfamily{#3}\fontseries{#4}\fontshape{#5}%
  \selectfont}%
\fi\endgroup%
\begin{picture}(398,398)(2097,-935)
\put(2206,-826){\makebox(0,0)[lb]{\smash{{\SetFigFont{20}{24.0}{\familydefault}{\mddefault}{\updefault}{\color[rgb]{0,0,0}1}%
}}}}
\end{picture}%
}}
\newcommand{\Atwo}[0]{\scalebox{0.4}{\begin{picture}(0,0)%
\includegraphics{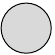}%
\end{picture}%
\setlength{\unitlength}{4144sp}%
\begingroup\makeatletter\ifx\SetFigFont\undefined%
\gdef\SetFigFont#1#2#3#4#5{%
  \reset@font\fontsize{#1}{#2pt}%
  \fontfamily{#3}\fontseries{#4}\fontshape{#5}%
  \selectfont}%
\fi\endgroup%
\begin{picture}(398,398)(2097,-935)
\put(2206,-826){\makebox(0,0)[lb]{\smash{{\SetFigFont{20}{24.0}{\familydefault}{\mddefault}{\updefault}{\color[rgb]{0,0,0}2}%
}}}}
\end{picture}%
}}
\newcommand{\Athree}[0]{\scalebox{0.4}{\begin{picture}(0,0)%
\includegraphics{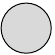}%
\end{picture}%
\setlength{\unitlength}{4144sp}%
\begingroup\makeatletter\ifx\SetFigFont\undefined%
\gdef\SetFigFont#1#2#3#4#5{%
  \reset@font\fontsize{#1}{#2pt}%
  \fontfamily{#3}\fontseries{#4}\fontshape{#5}%
  \selectfont}%
\fi\endgroup%
\begin{picture}(398,398)(2097,-935)
\put(2206,-826){\makebox(0,0)[lb]{\smash{{\SetFigFont{20}{24.0}{\familydefault}{\mddefault}{\updefault}{\color[rgb]{0,0,0}3}%
}}}}
\end{picture}%
}}
\newcommand{\Afour}[0]{\scalebox{0.4}{\begin{picture}(0,0)%
\includegraphics{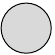}%
\end{picture}%
\setlength{\unitlength}{4144sp}%
\begingroup\makeatletter\ifx\SetFigFont\undefined%
\gdef\SetFigFont#1#2#3#4#5{%
  \reset@font\fontsize{#1}{#2pt}%
  \fontfamily{#3}\fontseries{#4}\fontshape{#5}%
  \selectfont}%
\fi\endgroup%
\begin{picture}(398,398)(2097,-935)
\put(2206,-826){\makebox(0,0)[lb]{\smash{{\SetFigFont{20}{24.0}{\familydefault}{\mddefault}{\updefault}{\color[rgb]{0,0,0}4}%
}}}}
\end{picture}%
}}
\newcommand{\Afive}[0]{\scalebox{0.4}{\begin{picture}(0,0)%
\includegraphics{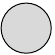}%
\end{picture}%
\setlength{\unitlength}{4144sp}%
\begingroup\makeatletter\ifx\SetFigFont\undefined%
\gdef\SetFigFont#1#2#3#4#5{%
  \reset@font\fontsize{#1}{#2pt}%
  \fontfamily{#3}\fontseries{#4}\fontshape{#5}%
  \selectfont}%
\fi\endgroup%
\begin{picture}(398,398)(2097,-935)
\put(2206,-826){\makebox(0,0)[lb]{\smash{{\SetFigFont{20}{24.0}{\familydefault}{\mddefault}{\updefault}{\color[rgb]{0,0,0}5}%
}}}}
\end{picture}%
}}
\newcommand{\Asix}[0]{\scalebox{0.4}{\begin{picture}(0,0)%
\includegraphics{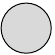}%
\end{picture}%
\setlength{\unitlength}{4144sp}%
\begingroup\makeatletter\ifx\SetFigFont\undefined%
\gdef\SetFigFont#1#2#3#4#5{%
  \reset@font\fontsize{#1}{#2pt}%
  \fontfamily{#3}\fontseries{#4}\fontshape{#5}%
  \selectfont}%
\fi\endgroup%
\begin{picture}(398,398)(2097,-935)
\put(2206,-826){\makebox(0,0)[lb]{\smash{{\SetFigFont{20}{24.0}{\familydefault}{\mddefault}{\updefault}{\color[rgb]{0,0,0}6}%
}}}}
\end{picture}%
}}
\def\N{{\ensuremath {\cal N}}}

\def\B{\ensuremath{\mathbb{B}}}
\def\Bn{\ensuremath{\B^n}}

\newcommand{\set}[1]{\ensuremath{\{#1\}}}

\newcommand{\litemlike}[1]{{\color{black!60}(#1)}}
\newcommand{\litemunlike}[1]{{\color{black!60}#1}}

\newcounter{listitem}
\newcounter{listitemr}
\renewcommand{\thelistitemr}{(\roman{listitemr})~}
\newcommand{\litemr}{{\color{black!60}\thelistitemr}\refstepcounter{listitemr}}

\newcounter{listitems}
\newcounter{listitemM}
\newcounter{listitema}
\renewcommand{\thelistitema}{(\arabic{listitema})~}
\newcommand{\litema}{{\color{black!60}\thelistitema}\refstepcounter{listitema}}

\def\endlitema{\setcounter{listitema}{1}}

\def\endlitemr{\setcounter{listitemr}{1}}

\newcounter{observation}
\newcounter{observationG}

\newenvironment{observation}{\refstepcounter{observation}{\color{gray}\sc Observation \theobservation:}\par\itshape}{\normalfont\bigskip}

\newenvironment{observationG}{\refstepcounter{observationG}{\color{gray}\sc Observation \theobservationG:}\par\itshape\color{gray}}{\normalfont\bigskip\color{black}}

\begin{document}

\title{{\Large Perspectives  and Networks}} %on Interaction Systems Boolean Automata 
\author{{\large Mathilde Noual}}
\maketitle

% +++++++++++++++++++++++++++++++++++++++++++++++++++++++++++++++++++++++++++++++
% +                                                                             +
% +++++++++++++++++++++++++++++++++++++++++++++++++++++++++++++++++++++++++++++++

%\section*{Introduction}

The {\em perspective} we take on a system determines the features and properties
of the system that we focus on. It determines where we search for causes to
explain the effects on the system that we observe. It determines the terms in
which we expect the information about the system to be expressed. And it can
also influence the choice of formalism that will be used to convey the
information. This paper proposes to start making these considerations concrete
in order to draw a practical benefit out of them.

% °°°°°°°°°°°°°°°°°°°°°°°°°°°°°°°°°°°°°°°°°°°°°°°°°°°°°°°°°°°°°°°°°°°°°°°°°°°°°°°
% °                                                                             °
% °°°°°°°°°°°°°°°°°°°°°°°°°°°°°°°°°°°°°°°°°°°°°°°°°°°°°°°°°°°°°°°°°°°°°°°°°°°°°°°

\section{Networks}
In the context of this paper, a {\em network} -- a.k.a.  an {\em
  interaction system} --  is any set of entities/parameters that we consider
as a whole for the following reason: we presume that the changes underwent by
the entities in the set are causally related to one another and account for the
global system changes that we are interested in.  
\bigskip

\colorbox{yellow!20}{\parbox{0.99\linewidth}{\begin{spacing}{1}\small {Remark:}
      {\it An interaction system can be comprised of just one entity (the human
        body for instance, as it might been seen in the light of a different
        culture). In this case, to explain the system's global state changes,
        the only thing we have is, precisely, the system's global state changes
        -- and also the system's environment, \ie{} everything that is not the
        system.  So to explain the system's global state changes by something
        else than themselves, we must turn to the environment and find an
        external cause. In other terms, we must add to the system a second
        entity (or more) that interacts with the original.  Under different
        circumstances, the same object might also be regarded as an interaction
        system comprised of {\em several} interacting entities (\eg the human
        body as seen by modern medicine). In this case, the environment is not
        the only place where explanations can be sought. Finer explanations
        might also come from considering alternative interactions between the
        different entities of the system. Gladly, since our present
        representations of objects can be questioned, in neither case are we
        condemned to uncovering the same sorts of explanations in the same sorts
        of ways. }
\end{spacing}}}
\bigskip

In this paper, we will illustrate the five following observations about
interaction systems.\bigskip

\begin{observation}
Some properties that we regard as {properties of 
systems} -- properties that the systems may have or not have -- are actually
{properties of the way we look at the systems}.
\label{obs:prop}
\end{observation}

\begin{observation}
A statement like ``{\em There exists no interaction between entity X and entity
Y.}''  has no essential meaning in itself. Its truth value is dependent on the
specific level of abstraction from which the system -- its entities, the
interactions between them -- is being looked at and defined.
\label{obs:exist}
\end{observation}

\begin{observation}
Incomplete data is not the only reason for the model of a real interaction 
system to fail to account for existing interactions between entities of the
system. Despite the modeller's flawless observation and formalisation of the
system, some causal relationships between entities might still be intrinsically
imperceptible to him/her under his/her current perspective.
\label{obs:deficit}
\end{observation}

\begin{observation}
 What information a formal object can provide about the real system it is meant
 to model, and what uninterpretable, non-modelling information it provides on
 top of that, does not just depend on the semantics associated to the formalism
 describing the formal object, it also essentially depends on the relative
 consistency of those semantics.
\label{obs:MT}
\end{observation}

\begin{observation}
The semantics associated to a formal object can be decisively affected by the
history of the object and how our scientific community came to inherit it. And
thus, so can the precise definition of the object that we choose among different
customary variations of the definition, and  the properties it has that we
take interest in, and those we don't.
\label{obs:com}
\end{observation}

% °°°°°°°°°°°°°°°°°°°°°°°°°°°°°°°°°°°°°°°°°°°°°°°°°°°°°°°°°°°°°°°°°°°°°°°°°°°°°°°
% °                                                                             °
% °°°°°°°°°°°°°°°°°°°°°°°°°°°°°°°°°°°°°°°°°°°°°°°°°°°°°°°°°°°°°°°°°°°°°°°°°°°°°°°

\section{Prototypes of networks}

To illustrate those five observations, we are going to use a
minimalist mathematical prototype of interaction systems named {\em Boolean
  Automata Networks} (BANs).\bigskip

\begin{figure}[h!]
\begin{tabular}{c@{\hspace{15mm}}c}
\fcolorbox{black}{black!5}{\scalebox{0.65}{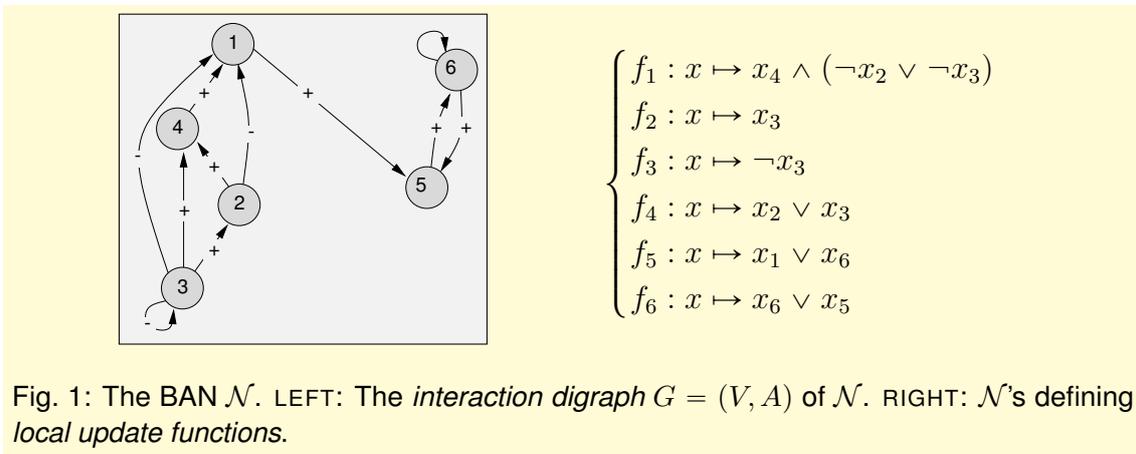}} \\[-40mm]
& $\begin{cases}
f_1:x\mapsto  x_4\wedge (\neg x_2\vee \neg x_3)\\
f_2:x\mapsto  x_3\\
f_3:x\mapsto  \neg x_3\\
f_4:x\mapsto  x_2\vee x_3\\
f_{5}:x\mapsto  x_1\vee x_{6}\\
f_{6}:x\mapsto x_{6}\vee x_{5}
\end{cases}$\\[20mm]\end{tabular}
\caption{The BAN $\N$. {\sc left:} The {\em interaction digraph} 
$G=(V,A)$ of $\N$. {\sc right:} $\N$'s defining {\em local update functions}.}
\label{fig:BANa}
\end{figure}

I will introduce the formalism of BANs using the BAN $\N$ represented below
in \figref{fig:BANa}. 
\bigskip

$\N$ is comprised of $n=6$ entities, namely \scalebox{0.4}{\begin{picture}(0,0)%
\includegraphics{automaton1.pdf}%
\end{picture}%
\setlength{\unitlength}{4144sp}%
\begingroup\makeatletter\ifx\SetFigFont\undefined%
\gdef\SetFigFont#1#2#3#4#5{%
  \reset@font\fontsize{#1}{#2pt}%
  \fontfamily{#3}\fontseries{#4}\fontshape{#5}%
  \selectfont}%
\fi\endgroup%
\begin{picture}(398,398)(2097,-935)
\put(2206,-826){\makebox(0,0)[lb]{\smash{{\SetFigFont{20}{24.0}{\familydefault}{\mddefault}{\updefault}{\color[rgb]{0,0,0}1}%
}}}}
\end{picture}%
}, \scalebox{0.4}{\begin{picture}(0,0)%
\includegraphics{automaton2.pdf}%
\end{picture}%
\setlength{\unitlength}{4144sp}%
\begingroup\makeatletter\ifx\SetFigFont\undefined%
\gdef\SetFigFont#1#2#3#4#5{%
  \reset@font\fontsize{#1}{#2pt}%
  \fontfamily{#3}\fontseries{#4}\fontshape{#5}%
  \selectfont}%
\fi\endgroup%
\begin{picture}(398,398)(2097,-935)
\put(2206,-826){\makebox(0,0)[lb]{\smash{{\SetFigFont{20}{24.0}{\familydefault}{\mddefault}{\updefault}{\color[rgb]{0,0,0}2}%
}}}}
\end{picture}%
}, \scalebox{0.4}{\begin{picture}(0,0)%
\includegraphics{automaton3.pdf}%
\end{picture}%
\setlength{\unitlength}{4144sp}%
\begingroup\makeatletter\ifx\SetFigFont\undefined%
\gdef\SetFigFont#1#2#3#4#5{%
  \reset@font\fontsize{#1}{#2pt}%
  \fontfamily{#3}\fontseries{#4}\fontshape{#5}%
  \selectfont}%
\fi\endgroup%
\begin{picture}(398,398)(2097,-935)
\put(2206,-826){\makebox(0,0)[lb]{\smash{{\SetFigFont{20}{24.0}{\familydefault}{\mddefault}{\updefault}{\color[rgb]{0,0,0}3}%
}}}}
\end{picture}%
}, \scalebox{0.4}{\begin{picture}(0,0)%
\includegraphics{automaton4.pdf}%
\end{picture}%
\setlength{\unitlength}{4144sp}%
\begingroup\makeatletter\ifx\SetFigFont\undefined%
\gdef\SetFigFont#1#2#3#4#5{%
  \reset@font\fontsize{#1}{#2pt}%
  \fontfamily{#3}\fontseries{#4}\fontshape{#5}%
  \selectfont}%
\fi\endgroup%
\begin{picture}(398,398)(2097,-935)
\put(2206,-826){\makebox(0,0)[lb]{\smash{{\SetFigFont{20}{24.0}{\familydefault}{\mddefault}{\updefault}{\color[rgb]{0,0,0}4}%
}}}}
\end{picture}%
},
\scalebox{0.4}{\begin{picture}(0,0)%
\includegraphics{automaton5.pdf}%
\end{picture}%
\setlength{\unitlength}{4144sp}%
\begingroup\makeatletter\ifx\SetFigFont\undefined%
\gdef\SetFigFont#1#2#3#4#5{%
  \reset@font\fontsize{#1}{#2pt}%
  \fontfamily{#3}\fontseries{#4}\fontshape{#5}%
  \selectfont}%
\fi\endgroup%
\begin{picture}(398,398)(2097,-935)
\put(2206,-826){\makebox(0,0)[lb]{\smash{{\SetFigFont{20}{24.0}{\familydefault}{\mddefault}{\updefault}{\color[rgb]{0,0,0}5}%
}}}}
\end{picture}%
}, and \scalebox{0.4}{\begin{picture}(0,0)%
\includegraphics{automaton6.pdf}%
\end{picture}%
\setlength{\unitlength}{4144sp}%
\begingroup\makeatletter\ifx\SetFigFont\undefined%
\gdef\SetFigFont#1#2#3#4#5{%
  \reset@font\fontsize{#1}{#2pt}%
  \fontfamily{#3}\fontseries{#4}\fontshape{#5}%
  \selectfont}%
\fi\endgroup%
\begin{picture}(398,398)(2097,-935)
\put(2206,-826){\makebox(0,0)[lb]{\smash{{\SetFigFont{20}{24.0}{\familydefault}{\mddefault}{\updefault}{\color[rgb]{0,0,0}6}%
}}}}
\end{picture}%
}\,, a.k.a. {automata} $1,2,3,4,5$ and $6$.
Each automaton $i\in V=\llbracket 1,n\rrbracket$ has a variable state in
$\B=\set{0,1}$.  If $\forall i\in V,\, x_i$ denotes the current state of $i$,
then $x=(x_1,x_2,\ldots,x_n)\in \Bn$ denotes the current state of $\N$.  On the
right of \figref{fig:BANa} are listed six functions \mbox{$f_i:\Bn\to \B$}. One
for each automaton  \scalebox{0.4}{\begin{picture}(0,0)%
\includegraphics{automatoni.pdf}%
\end{picture}%
\setlength{\unitlength}{4144sp}%
\begingroup\makeatletter\ifx\SetFigFont\undefined%
\gdef\SetFigFont#1#2#3#4#5{%
  \reset@font\fontsize{#1}{#2pt}%
  \fontfamily{#3}\fontseries{#4}\fontshape{#5}%
  \selectfont}%
\fi\endgroup%
\begin{picture}(398,398)(2097,-935)
\put(2206,-826){\makebox(0,0)[lb]{\smash{{\SetFigFont{20}{24.0}{\familydefault}{\mddefault}{\updefault}{\color[rgb]{0,0,0}$i$}%
}}}}
\end{picture}%
}{}.  Function $f_i$ defines the possible variations of \scalebox{0.4}{\begin{picture}(0,0)%
\includegraphics{automatoni.pdf}%
\end{picture}%
\setlength{\unitlength}{4144sp}%
\begingroup\makeatletter\ifx\SetFigFont\undefined%
\gdef\SetFigFont#1#2#3#4#5{%
  \reset@font\fontsize{#1}{#2pt}%
  \fontfamily{#3}\fontseries{#4}\fontshape{#5}%
  \selectfont}%
\fi\endgroup%
\begin{picture}(398,398)(2097,-935)
\put(2206,-826){\makebox(0,0)[lb]{\smash{{\SetFigFont{20}{24.0}{\familydefault}{\mddefault}{\updefault}{\color[rgb]{0,0,0}$i$}%
}}}}
\end{picture}%
}{}'s
state depending on $\N$'s.
In state $x\in \Bn$ of $\N$, \scalebox{0.4}{\begin{picture}(0,0)%
\includegraphics{automatoni.pdf}%
\end{picture}%
\setlength{\unitlength}{4144sp}%
\begingroup\makeatletter\ifx\SetFigFont\undefined%
\gdef\SetFigFont#1#2#3#4#5{%
  \reset@font\fontsize{#1}{#2pt}%
  \fontfamily{#3}\fontseries{#4}\fontshape{#5}%
  \selectfont}%
\fi\endgroup%
\begin{picture}(398,398)(2097,-935)
\put(2206,-826){\makebox(0,0)[lb]{\smash{{\SetFigFont{20}{24.0}{\familydefault}{\mddefault}{\updefault}{\color[rgb]{0,0,0}$i$}%
}}}}
\end{picture}%
}{}'s state \underline{can} change if and
only if \scalebox{0.4}{\begin{picture}(0,0)%
\includegraphics{automatoni.pdf}%
\end{picture}%
\setlength{\unitlength}{4144sp}%
\begingroup\makeatletter\ifx\SetFigFont\undefined%
\gdef\SetFigFont#1#2#3#4#5{%
  \reset@font\fontsize{#1}{#2pt}%
  \fontfamily{#3}\fontseries{#4}\fontshape{#5}%
  \selectfont}%
\fi\endgroup%
\begin{picture}(398,398)(2097,-935)
\put(2206,-826){\makebox(0,0)[lb]{\smash{{\SetFigFont{20}{24.0}{\familydefault}{\mddefault}{\updefault}{\color[rgb]{0,0,0}$i$}%
}}}}
\end{picture}%
}{} belongs to the set $U(x)=\set{i\in V: f_i(x)\neq x_i}$ of automata
that are {unstable} in $x$.  The state of \scalebox{0.4}{\begin{picture}(0,0)%
\includegraphics{automatoni.pdf}%
\end{picture}%
\setlength{\unitlength}{4144sp}%
\begingroup\makeatletter\ifx\SetFigFont\undefined%
\gdef\SetFigFont#1#2#3#4#5{%
  \reset@font\fontsize{#1}{#2pt}%
  \fontfamily{#3}\fontseries{#4}\fontshape{#5}%
  \selectfont}%
\fi\endgroup%
\begin{picture}(398,398)(2097,-935)
\put(2206,-826){\makebox(0,0)[lb]{\smash{{\SetFigFont{20}{24.0}{\familydefault}{\mddefault}{\updefault}{\color[rgb]{0,0,0}$i$}%
}}}}
\end{picture}%
}{} actually \underline{\em does}
change if: \litemr it \underline{\em can} and \litemr\endlitemr \scalebox{0.4}{\begin{picture}(0,0)%
\includegraphics{automatoni.pdf}%
\end{picture}%
\setlength{\unitlength}{4144sp}%
\begingroup\makeatletter\ifx\SetFigFont\undefined%
\gdef\SetFigFont#1#2#3#4#5{%
  \reset@font\fontsize{#1}{#2pt}%
  \fontfamily{#3}\fontseries{#4}\fontshape{#5}%
  \selectfont}%
\fi\endgroup%
\begin{picture}(398,398)(2097,-935)
\put(2206,-826){\makebox(0,0)[lb]{\smash{{\SetFigFont{20}{24.0}{\familydefault}{\mddefault}{\updefault}{\color[rgb]{0,0,0}$i$}%
}}}}
\end{picture}%
}{} is
updated.  For instance, in state $x=(1,1,1,0,0,1)$ of $\N$, if \scalebox{0.4}{\begin{picture}(0,0)%
\includegraphics{automaton1.pdf}%
\end{picture}%
\setlength{\unitlength}{4144sp}%
\begingroup\makeatletter\ifx\SetFigFont\undefined%
\gdef\SetFigFont#1#2#3#4#5{%
  \reset@font\fontsize{#1}{#2pt}%
  \fontfamily{#3}\fontseries{#4}\fontshape{#5}%
  \selectfont}%
\fi\endgroup%
\begin{picture}(398,398)(2097,-935)
\put(2206,-826){\makebox(0,0)[lb]{\smash{{\SetFigFont{20}{24.0}{\familydefault}{\mddefault}{\updefault}{\color[rgb]{0,0,0}1}%
}}}}
\end{picture}%
}, \scalebox{0.4}{\begin{picture}(0,0)%
\includegraphics{automaton2.pdf}%
\end{picture}%
\setlength{\unitlength}{4144sp}%
\begingroup\makeatletter\ifx\SetFigFont\undefined%
\gdef\SetFigFont#1#2#3#4#5{%
  \reset@font\fontsize{#1}{#2pt}%
  \fontfamily{#3}\fontseries{#4}\fontshape{#5}%
  \selectfont}%
\fi\endgroup%
\begin{picture}(398,398)(2097,-935)
\put(2206,-826){\makebox(0,0)[lb]{\smash{{\SetFigFont{20}{24.0}{\familydefault}{\mddefault}{\updefault}{\color[rgb]{0,0,0}2}%
}}}}
\end{picture}%
}, and
\scalebox{0.4}{\begin{picture}(0,0)%
\includegraphics{automaton4.pdf}%
\end{picture}%
\setlength{\unitlength}{4144sp}%
\begingroup\makeatletter\ifx\SetFigFont\undefined%
\gdef\SetFigFont#1#2#3#4#5{%
  \reset@font\fontsize{#1}{#2pt}%
  \fontfamily{#3}\fontseries{#4}\fontshape{#5}%
  \selectfont}%
\fi\endgroup%
\begin{picture}(398,398)(2097,-935)
\put(2206,-826){\makebox(0,0)[lb]{\smash{{\SetFigFont{20}{24.0}{\familydefault}{\mddefault}{\updefault}{\color[rgb]{0,0,0}4}%
}}}}
\end{picture}%
}{} are updated, and \scalebox{0.4}{\begin{picture}(0,0)%
\includegraphics{automaton3.pdf}%
\end{picture}%
\setlength{\unitlength}{4144sp}%
\begingroup\makeatletter\ifx\SetFigFont\undefined%
\gdef\SetFigFont#1#2#3#4#5{%
  \reset@font\fontsize{#1}{#2pt}%
  \fontfamily{#3}\fontseries{#4}\fontshape{#5}%
  \selectfont}%
\fi\endgroup%
\begin{picture}(398,398)(2097,-935)
\put(2206,-826){\makebox(0,0)[lb]{\smash{{\SetFigFont{20}{24.0}{\familydefault}{\mddefault}{\updefault}{\color[rgb]{0,0,0}3}%
}}}}
\end{picture}%
}, \scalebox{0.4}{\begin{picture}(0,0)%
\includegraphics{automaton5.pdf}%
\end{picture}%
\setlength{\unitlength}{4144sp}%
\begingroup\makeatletter\ifx\SetFigFont\undefined%
\gdef\SetFigFont#1#2#3#4#5{%
  \reset@font\fontsize{#1}{#2pt}%
  \fontfamily{#3}\fontseries{#4}\fontshape{#5}%
  \selectfont}%
\fi\endgroup%
\begin{picture}(398,398)(2097,-935)
\put(2206,-826){\makebox(0,0)[lb]{\smash{{\SetFigFont{20}{24.0}{\familydefault}{\mddefault}{\updefault}{\color[rgb]{0,0,0}5}%
}}}}
\end{picture}%
}, and \scalebox{0.4}{\begin{picture}(0,0)%
\includegraphics{automaton6.pdf}%
\end{picture}%
\setlength{\unitlength}{4144sp}%
\begingroup\makeatletter\ifx\SetFigFont\undefined%
\gdef\SetFigFont#1#2#3#4#5{%
  \reset@font\fontsize{#1}{#2pt}%
  \fontfamily{#3}\fontseries{#4}\fontshape{#5}%
  \selectfont}%
\fi\endgroup%
\begin{picture}(398,398)(2097,-935)
\put(2206,-826){\makebox(0,0)[lb]{\smash{{\SetFigFont{20}{24.0}{\familydefault}{\mddefault}{\updefault}{\color[rgb]{0,0,0}6}%
}}}}
\end{picture}%
}{} aren't, then $\N$ transitions
to state $x'=(f_1(x),f_2(x),x_3,f_4(x),x_5,x_6)=(0,1,1,1,0,1)$.\bigskip

Formally, a BAN is defined as a set of Boolean functions
$\N=\set{f_i:\Bn\to \B\,|\,i\in V}$ with no mention to automata updates. A BAN is
therefore {\em not} a dynamical system.  This choice of definition is deliberate
although nontraditional.  In the literature, Automata Networks (ANs) are usually
taken to represent dynamical
systems \citemycite{Barrett2001,ColonReyes2004,Delaplace2011,Demongeot2000,Goles1990B,Goles2008,Jarrah2010,ConfuseParallelAndSynchronism6Naldi,Mortveit2001,MortveitCalculus,Aracena2009,Mendoza1998,Remy2008a,Remy2008b,Remy2003,Richard2007,Richard2009,Robert1986B,Siebert2009,Tournier2009}{DAM2010,AAM2011,BMB2013,DAM2011,PLOS2010,IJMS2009,JTB2011,Demongeot2008c}.
Their definitions are made to imply specific updating conditions. The notions of
causality and time are thereby fused together and a distinction is implied
between \litema process of change and
\litema\endlitema (result of) change -- the latter being either an approximation
of the former or an observed consequence of it.
The object of this paper requires to take a less abstract view on the formalism,
and not make this semantical distinction, so: \linebreak\vspace{-4mm}

{\em Change is anything that has
observable consequence, if only the observable consequence of us noticing
it. Process of change is change if it has observable consequence {in itself};
if it hasn't, then it's not something there is anything to be said about. }\medskip

 In
agreement with the formalism of BANs, we will assume that in a BAN, all considered
changes are represented with the same status. The focus is therefore not so much
on automata {states}, as it is on their {changes}.

\bigskip\bigskip

\section{Illustrations}

% °°°°°°°°°°°°°°°°°°°°°°°°°°°°°°°°°°°°°°°°°°°°°°°°°°°°°°°°°°°°°°°°°°°°°°°°°°°°°°°
% °                                                                             °
% °°°°°°°°°°°°°°°°°°°°°°°°°°°°°°°°°°°°°°°°°°°°°°°°°°°°°°°°°°°°°°°°°°°°°°°°°°°°°°°
\begin{observationG}
Some properties that we regard as {properties of interaction
systems} -- properties that they can have or not have -- are actually
{properties of the way we look at them}.
\label{obs:propG}
\end{observationG}

% °°°°°°°°°°°°°°°°°°°°°°°°°°°°°°°°°°°°°°°°°°°°°°°°°°°°°°°°°°°°°°°°°°°°°°°°°°°°°°°
% °                                                                             °
% °°°°°°°°°°°°°°°°°°°°°°°°°°°°°°°°°°°°°°°°°°°°°°°°°°°°°°°°°°°°°°°°°°°°°°°°°°°°°°°

Observation 1 can be illustrated with the BAN properties of {\em monotony} and
{\em non-monotony}. A BAN is said to be monotone when the following holds
for any two of its automata \scalebox{0.4}{\begin{picture}(0,0)%
\includegraphics{automatoni.pdf}%
\end{picture}%
\setlength{\unitlength}{4144sp}%
\begingroup\makeatletter\ifx\SetFigFont\undefined%
\gdef\SetFigFont#1#2#3#4#5{%
  \reset@font\fontsize{#1}{#2pt}%
  \fontfamily{#3}\fontseries{#4}\fontshape{#5}%
  \selectfont}%
\fi\endgroup%
\begin{picture}(398,398)(2097,-935)
\put(2206,-826){\makebox(0,0)[lb]{\smash{{\SetFigFont{20}{24.0}{\familydefault}{\mddefault}{\updefault}{\color[rgb]{0,0,0}$i$}%
}}}}
\end{picture}%
}{} and \scalebox{0.4}{\begin{picture}(0,0)%
\includegraphics{automatonj.pdf}%
\end{picture}%
\setlength{\unitlength}{4144sp}%
\begingroup\makeatletter\ifx\SetFigFont\undefined%
\gdef\SetFigFont#1#2#3#4#5{%
  \reset@font\fontsize{#1}{#2pt}%
  \fontfamily{#3}\fontseries{#4}\fontshape{#5}%
  \selectfont}%
\fi\endgroup%
\begin{picture}(398,398)(2097,-935)
\put(2206,-826){\makebox(0,0)[lb]{\smash{{\SetFigFont{20}{24.0}{\familydefault}{\mddefault}{\updefault}{\color[rgb]{0,0,0}$j$}%
}}}}
\end{picture}%
}. If \scalebox{0.4}{\begin{picture}(0,0)%
\includegraphics{automatoni.pdf}%
\end{picture}%
\setlength{\unitlength}{4144sp}%
\begingroup\makeatletter\ifx\SetFigFont\undefined%
\gdef\SetFigFont#1#2#3#4#5{%
  \reset@font\fontsize{#1}{#2pt}%
  \fontfamily{#3}\fontseries{#4}\fontshape{#5}%
  \selectfont}%
\fi\endgroup%
\begin{picture}(398,398)(2097,-935)
\put(2206,-826){\makebox(0,0)[lb]{\smash{{\SetFigFont{20}{24.0}{\familydefault}{\mddefault}{\updefault}{\color[rgb]{0,0,0}$i$}%
}}}}
\end{picture}%
}{} influences \scalebox{0.4}{\begin{picture}(0,0)%
\includegraphics{automatonj.pdf}%
\end{picture}%
\setlength{\unitlength}{4144sp}%
\begingroup\makeatletter\ifx\SetFigFont\undefined%
\gdef\SetFigFont#1#2#3#4#5{%
  \reset@font\fontsize{#1}{#2pt}%
  \fontfamily{#3}\fontseries{#4}\fontshape{#5}%
  \selectfont}%
\fi\endgroup%
\begin{picture}(398,398)(2097,-935)
\put(2206,-826){\makebox(0,0)[lb]{\smash{{\SetFigFont{20}{24.0}{\familydefault}{\mddefault}{\updefault}{\color[rgb]{0,0,0}$j$}%
}}}}
\end{picture}%
}{}, then it always
does so in the same way: either  \scalebox{0.4}{\begin{picture}(0,0)%
\includegraphics{automatoni.pdf}%
\end{picture}%
\setlength{\unitlength}{4144sp}%
\begingroup\makeatletter\ifx\SetFigFont\undefined%
\gdef\SetFigFont#1#2#3#4#5{%
  \reset@font\fontsize{#1}{#2pt}%
  \fontfamily{#3}\fontseries{#4}\fontshape{#5}%
  \selectfont}%
\fi\endgroup%
\begin{picture}(398,398)(2097,-935)
\put(2206,-826){\makebox(0,0)[lb]{\smash{{\SetFigFont{20}{24.0}{\familydefault}{\mddefault}{\updefault}{\color[rgb]{0,0,0}$i$}%
}}}}
\end{picture}%
}{} always influences \scalebox{0.4}{\begin{picture}(0,0)%
\includegraphics{automatonj.pdf}%
\end{picture}%
\setlength{\unitlength}{4144sp}%
\begingroup\makeatletter\ifx\SetFigFont\undefined%
\gdef\SetFigFont#1#2#3#4#5{%
  \reset@font\fontsize{#1}{#2pt}%
  \fontfamily{#3}\fontseries{#4}\fontshape{#5}%
  \selectfont}%
\fi\endgroup%
\begin{picture}(398,398)(2097,-935)
\put(2206,-826){\makebox(0,0)[lb]{\smash{{\SetFigFont{20}{24.0}{\familydefault}{\mddefault}{\updefault}{\color[rgb]{0,0,0}$j$}%
}}}}
\end{picture}%
}{} positively, or
\scalebox{0.4}{\begin{picture}(0,0)%
\includegraphics{automatoni.pdf}%
\end{picture}%
\setlength{\unitlength}{4144sp}%
\begingroup\makeatletter\ifx\SetFigFont\undefined%
\gdef\SetFigFont#1#2#3#4#5{%
  \reset@font\fontsize{#1}{#2pt}%
  \fontfamily{#3}\fontseries{#4}\fontshape{#5}%
  \selectfont}%
\fi\endgroup%
\begin{picture}(398,398)(2097,-935)
\put(2206,-826){\makebox(0,0)[lb]{\smash{{\SetFigFont{20}{24.0}{\familydefault}{\mddefault}{\updefault}{\color[rgb]{0,0,0}$i$}%
}}}}
\end{picture}%
}{} always influences \scalebox{0.4}{\begin{picture}(0,0)%
\includegraphics{automatonj.pdf}%
\end{picture}%
\setlength{\unitlength}{4144sp}%
\begingroup\makeatletter\ifx\SetFigFont\undefined%
\gdef\SetFigFont#1#2#3#4#5{%
  \reset@font\fontsize{#1}{#2pt}%
  \fontfamily{#3}\fontseries{#4}\fontshape{#5}%
  \selectfont}%
\fi\endgroup%
\begin{picture}(398,398)(2097,-935)
\put(2206,-826){\makebox(0,0)[lb]{\smash{{\SetFigFont{20}{24.0}{\familydefault}{\mddefault}{\updefault}{\color[rgb]{0,0,0}$j$}%
}}}}
\end{picture}%
}{} negatively.
In mathematical terms this translates to the following where $\bar{x}^{i}\in
\Bn$ is the state defined by $\forall k\neq i: \bar{x}^{i}_k=x_k$ and
$\bar{x}^{i}=\neg x_i$.  By definition, in a monotone BAN, for any $i,j\in V$
such that $(i,j)\in A$, \ie such that the Conjunctive Normal Form of $f_j(x)$
depends on $x_i$:\linebreak either $\forall x\in\Bn,\, x_i=1\Rightarrow f_j(x)\geq
f_j(\bar{x}^{i})$, or $\forall x\in\Bn,\, x_i=1\Rightarrow f_j(x)\leq
f_j(\bar{x}^{i})$.
\medskip

\begin{figure}[h!]
\begin{tabular}{c@{\hspace{10mm}}c}
\fcolorbox{black}{black!5}{\scalebox{0.65}{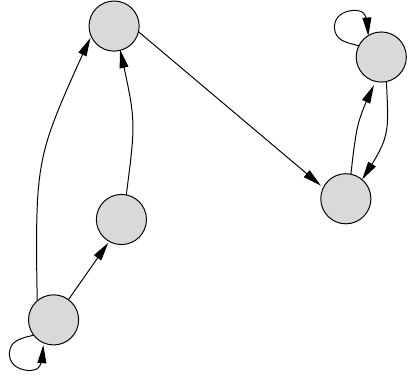}} \\[-40mm]
& $\begin{cases}
\widehat{f}_1:x\mapsto  x_2\oplus x_3=(\neg x_2\vee \neg x_3)\wedge (x_2\vee x_3)\\
\widehat{f}_2:x\mapsto  x_3\\
\widehat{f}_3:x\mapsto  \neg x_3\\
~\\
\widehat{f}_{5}:x\mapsto  x_1\vee x_{6}\\
\widehat{f}_{6}:x\mapsto x_{6}\vee x_{5}
\end{cases}$\\[20mm]\end{tabular}
\caption{The BAN $\widehat{\N}$. {\sc left:} The {interaction digraph} 
$\widehat{G}=(\widehat{V},\widehat{A})$ of $\widehat{\N}$. {\sc right:} $\widehat{\N}$'s defining local update functions.}
\label{fig:BANb}
\end{figure}

Traditionally, when BANs model biological (genetic) regulation
networks \citeal{Remy2003,Chaouiya2004,Aracena200649,Mendoza1998,Blanchini000562},
they are supposed to be monotone, like BAN $\N$ of \figref{fig:BANa} and unlike
BAN $\widehat{\N}$ of \figref{fig:BANb}. \bigskip
%\clearpage

Despite BAN $\widehat{\N}$ being  non-monotone, it is the exact
representation of what we see of the monotone BAN $\N$ in some
circumstances. \bigskip

Imagine that there exists a real system in nature that works exactly like the monotone
BAN $\N$ of \figref{fig:BANa} does.  Call the  real system $\N$ too. Imagine that we human
observers of reality are observing $\N$ in action, and at the time we are doing
that, for some reason, parts of $\N$ are behaving rhythmically: state changes of
\scalebox{0.4}{\begin{picture}(0,0)%
\includegraphics{automaton1.pdf}%
\end{picture}%
\setlength{\unitlength}{4144sp}%
\begingroup\makeatletter\ifx\SetFigFont\undefined%
\gdef\SetFigFont#1#2#3#4#5{%
  \reset@font\fontsize{#1}{#2pt}%
  \fontfamily{#3}\fontseries{#4}\fontshape{#5}%
  \selectfont}%
\fi\endgroup%
\begin{picture}(398,398)(2097,-935)
\put(2206,-826){\makebox(0,0)[lb]{\smash{{\SetFigFont{20}{24.0}{\familydefault}{\mddefault}{\updefault}{\color[rgb]{0,0,0}1}%
}}}}
\end{picture}%
}{} and \scalebox{0.4}{\begin{picture}(0,0)%
\includegraphics{automaton4.pdf}%
\end{picture}%
\setlength{\unitlength}{4144sp}%
\begingroup\makeatletter\ifx\SetFigFont\undefined%
\gdef\SetFigFont#1#2#3#4#5{%
  \reset@font\fontsize{#1}{#2pt}%
  \fontfamily{#3}\fontseries{#4}\fontshape{#5}%
  \selectfont}%
\fi\endgroup%
\begin{picture}(398,398)(2097,-935)
\put(2206,-826){\makebox(0,0)[lb]{\smash{{\SetFigFont{20}{24.0}{\familydefault}{\mddefault}{\updefault}{\color[rgb]{0,0,0}4}%
}}}}
\end{picture}%
}{} are happening at the same frequency  with a slight 
phase offset; everything is exactly as if \scalebox{0.4}{\begin{picture}(0,0)%
\includegraphics{automaton4.pdf}%
\end{picture}%
\setlength{\unitlength}{4144sp}%
\begingroup\makeatletter\ifx\SetFigFont\undefined%
\gdef\SetFigFont#1#2#3#4#5{%
  \reset@font\fontsize{#1}{#2pt}%
  \fontfamily{#3}\fontseries{#4}\fontshape{#5}%
  \selectfont}%
\fi\endgroup%
\begin{picture}(398,398)(2097,-935)
\put(2206,-826){\makebox(0,0)[lb]{\smash{{\SetFigFont{20}{24.0}{\familydefault}{\mddefault}{\updefault}{\color[rgb]{0,0,0}4}%
}}}}
\end{picture}%
}{} was systematically updated
immediately before \scalebox{0.4}{\begin{picture}(0,0)%
\includegraphics{automaton1.pdf}%
\end{picture}%
\setlength{\unitlength}{4144sp}%
\begingroup\makeatletter\ifx\SetFigFont\undefined%
\gdef\SetFigFont#1#2#3#4#5{%
  \reset@font\fontsize{#1}{#2pt}%
  \fontfamily{#3}\fontseries{#4}\fontshape{#5}%
  \selectfont}%
\fi\endgroup%
\begin{picture}(398,398)(2097,-935)
\put(2206,-826){\makebox(0,0)[lb]{\smash{{\SetFigFont{20}{24.0}{\familydefault}{\mddefault}{\updefault}{\color[rgb]{0,0,0}1}%
}}}}
\end{picture}%
}{}\, is.  If we were witnessing each event occurring in
$\N$ and knew that we were, then we would have enough information to build the
representation of $\N$ given in \figref{fig:BANa}. But assume that deliberately,
we are considering $\N$ from a specific level of abstraction (as opposed to
considering $\N$ from the godlike perspective of the Laplacian demon that sees
Everything because it is interested in Everything).  In other terms, assume we
have specific interests, and because of that we are focusing on specific
attributes of $\N$.  Imagine that in the present case, this results in us being
unaware of \scalebox{0.4}{\begin{picture}(0,0)%
\includegraphics{automaton4.pdf}%
\end{picture}%
\setlength{\unitlength}{4144sp}%
\begingroup\makeatletter\ifx\SetFigFont\undefined%
\gdef\SetFigFont#1#2#3#4#5{%
  \reset@font\fontsize{#1}{#2pt}%
  \fontfamily{#3}\fontseries{#4}\fontshape{#5}%
  \selectfont}%
\fi\endgroup%
\begin{picture}(398,398)(2097,-935)
\put(2206,-826){\makebox(0,0)[lb]{\smash{{\SetFigFont{20}{24.0}{\familydefault}{\mddefault}{\updefault}{\color[rgb]{0,0,0}4}%
}}}}
\end{picture}%
}{}'s existence. \bigskip 

{\sc nb:} This does not imply a default in our observation. What it is we are
looking at in the entities \scalebox{0.4}{\begin{picture}(0,0)%
\includegraphics{automatoni.pdf}%
\end{picture}%
\setlength{\unitlength}{4144sp}%
\begingroup\makeatletter\ifx\SetFigFont\undefined%
\gdef\SetFigFont#1#2#3#4#5{%
  \reset@font\fontsize{#1}{#2pt}%
  \fontfamily{#3}\fontseries{#4}\fontshape{#5}%
  \selectfont}%
\fi\endgroup%
\begin{picture}(398,398)(2097,-935)
\put(2206,-826){\makebox(0,0)[lb]{\smash{{\SetFigFont{20}{24.0}{\familydefault}{\mddefault}{\updefault}{\color[rgb]{0,0,0}$i$}%
}}}}
\end{picture}%
}\, $i\neq 4$ of system $\N$ might simply not
exist, not make sense, or not be measurable in entity \scalebox{0.4}{\begin{picture}(0,0)%
\includegraphics{automaton4.pdf}%
\end{picture}%
\setlength{\unitlength}{4144sp}%
\begingroup\makeatletter\ifx\SetFigFont\undefined%
\gdef\SetFigFont#1#2#3#4#5{%
  \reset@font\fontsize{#1}{#2pt}%
  \fontfamily{#3}\fontseries{#4}\fontshape{#5}%
  \selectfont}%
\fi\endgroup%
\begin{picture}(398,398)(2097,-935)
\put(2206,-826){\makebox(0,0)[lb]{\smash{{\SetFigFont{20}{24.0}{\familydefault}{\mddefault}{\updefault}{\color[rgb]{0,0,0}4}%
}}}}
\end{picture}%
}{}.  For
instance, \scalebox{0.4}{\begin{picture}(0,0)%
\includegraphics{automaton4.pdf}%
\end{picture}%
\setlength{\unitlength}{4144sp}%
\begingroup\makeatletter\ifx\SetFigFont\undefined%
\gdef\SetFigFont#1#2#3#4#5{%
  \reset@font\fontsize{#1}{#2pt}%
  \fontfamily{#3}\fontseries{#4}\fontshape{#5}%
  \selectfont}%
\fi\endgroup%
\begin{picture}(398,398)(2097,-935)
\put(2206,-826){\makebox(0,0)[lb]{\smash{{\SetFigFont{20}{24.0}{\familydefault}{\mddefault}{\updefault}{\color[rgb]{0,0,0}4}%
}}}}
\end{picture}%
}'s {state changes} might represent rapid decoding of mRNA
sequences, while the state changes of the other \scalebox{0.4}{\begin{picture}(0,0)%
\includegraphics{automatoni.pdf}%
\end{picture}%
\setlength{\unitlength}{4144sp}%
\begingroup\makeatletter\ifx\SetFigFont\undefined%
\gdef\SetFigFont#1#2#3#4#5{%
  \reset@font\fontsize{#1}{#2pt}%
  \fontfamily{#3}\fontseries{#4}\fontshape{#5}%
  \selectfont}%
\fi\endgroup%
\begin{picture}(398,398)(2097,-935)
\put(2206,-826){\makebox(0,0)[lb]{\smash{{\SetFigFont{20}{24.0}{\familydefault}{\mddefault}{\updefault}{\color[rgb]{0,0,0}$i$}%
}}}}
\end{picture}%
}{} $i\neq 4$ might represent
slower processes such the increase of protein concentrations in the cell during
the protein's synthesis.
\bigskip

Every time we witness \scalebox{0.4}{\begin{picture}(0,0)%
\includegraphics{automaton1.pdf}%
\end{picture}%
\setlength{\unitlength}{4144sp}%
\begingroup\makeatletter\ifx\SetFigFont\undefined%
\gdef\SetFigFont#1#2#3#4#5{%
  \reset@font\fontsize{#1}{#2pt}%
  \fontfamily{#3}\fontseries{#4}\fontshape{#5}%
  \selectfont}%
\fi\endgroup%
\begin{picture}(398,398)(2097,-935)
\put(2206,-826){\makebox(0,0)[lb]{\smash{{\SetFigFont{20}{24.0}{\familydefault}{\mddefault}{\updefault}{\color[rgb]{0,0,0}1}%
}}}}
\end{picture}%
}{} change states, \scalebox{0.4}{\begin{picture}(0,0)%
\includegraphics{automaton4.pdf}%
\end{picture}%
\setlength{\unitlength}{4144sp}%
\begingroup\makeatletter\ifx\SetFigFont\undefined%
\gdef\SetFigFont#1#2#3#4#5{%
  \reset@font\fontsize{#1}{#2pt}%
  \fontfamily{#3}\fontseries{#4}\fontshape{#5}%
  \selectfont}%
\fi\endgroup%
\begin{picture}(398,398)(2097,-935)
\put(2206,-826){\makebox(0,0)[lb]{\smash{{\SetFigFont{20}{24.0}{\familydefault}{\mddefault}{\updefault}{\color[rgb]{0,0,0}4}%
}}}}
\end{picture}%
}{} just has. While $\N$
is taking trajectory: 

\begin{gather*}\raisebox{1.8pt}{\ldots}\hspace{-4pt}\longrightarrow
x=(x_1,x_2,x_3,x_4,x_5,x_6)\stackrel{{\color{black!20}\scriptsize
4}}{\longrightarrow}
x'=(x_1,x_2,x_3,f_4(x),x_5,x_6)\\\hspace{5cm}\stackrel{{\color{black!20}\scriptsize
1}}{\longrightarrow}
x''=(f_1(x'),,x_2,x_3,f_4(x),x_5,x_6)\longrightarrow\hspace{-3pt}\raisebox{1.8pt}{\ldots}
\end{gather*}

we observers are just seeing:

\begin{gather*}
\raisebox{1.8pt}{\ldots}\hspace{-4pt}\longrightarrow
(x_1,x_2,x_3,x_5,x_6)\stackrel{{\color{black!20}\scriptsize 1}}{\longrightarrow}
(f_1(x'),x_2,x_3,x_5,x_6)\longrightarrow\hspace{-3pt}\raisebox{1.8pt}{\ldots}.
\end{gather*}

The BAN description of a system that behaves like this is the BAN description
given in \figref{fig:BANb}. Under such circumstances -- circumstances that
constrain the temporality of events in $\N$ together with the way we observe
those events, the level of abstraction from which we do that, and {\em the
  temporality of our observations of $\N$'s changes with respect to the
  temporality of $\N$'s changes} -- what is given of $\N$ for us to understand
is $\widehat{\N}$.\bigskip

\textls[-5]{In agreement with Observation 1, this shows that monotony and non-monotony are
not so much properties qualifying the interactions of a system as they are of
how  we look at it.}\linebreak

In the literature, wherever BANs are considered as stand-alone mathematical
objects, it is customary to restrict the local update functions $f_i$ to a
certain class of functions for convenience.  A typical example is the
restriction to functions that are expressible in terms of a limited number of
logical connectors
\citemycite{GolesOlivos81,ColonReyes2004,Mortveit2001,Laubenbacher2012B,MortveitElements,Aracena2004b,Jarrah2010,Goles1982a}{DAM2010,TCS2013,AAM2011,AUTOMATA2010}. And
as mentioned above, the $f_i$'s are also often restricted to functions that are
expressible, on the contrary, {\em without} certain connectors such as the
$\oplus$ (XOR) connector which makes the BAN severely {non-monotone}
\citemycite{MortveitSDS,Perrot2015}{AUTOMATA2012,TCS2013}.
\bigskip

If we want the mathematical understanding we develop about mathematical
representations of 'real' interaction systems to apply and to apply rightly,
then we need to understand the meaning of the restrictions we make when we
derive this mathematical understanding. Observation 1 shows how important
it is to consider thoroughly the way our perspective on a system and our
interpretation of its representation are involved  in
the properties that we build our understanding on.
\bigskip\bigskip\bigskip

% °°°°°°°°°°°°°°°°°°°°°°°°°°°°°°°°°°°°°°°°°°°°°°°°°°°°°°°°°°°°°°°°°°°°°°°°°°°°°°°
% °                                                                             °
% °°°°°°°°°°°°°°°°°°°°°°°°°°°°°°°°°°°°°°°°°°°°°°°°°°°°°°°°°°°°°°°°°°°°°°°°°°°°°°°
\begin{observationG}
A statement like ``{\em There exists no interaction between entity X and entity
Y.}''  has no essential meaning in itself. Its truth value is dependent on the
specific level of abstraction from which the system -- its entities, the
interactions between them -- is being looked at.
\label{obs:existG}
\end{observationG}

\begin{observationG}
Incomplete data is not the only reason for the model of a real interaction 
system to fail to account for existing interactions between entities of the
system. Despite the modeller's flawless observation and formalisation of the
system, some causal relationships between entities might still be intrinsically
imperceptible to him/her under his/her current perspective.
\label{obs:deficitG}
\end{observationG}

\begin{figure}[h!]
\begin{tabular}{c@{\hspace{15mm}}c}
\fcolorbox{black}{black!5}{\scalebox{0.65}{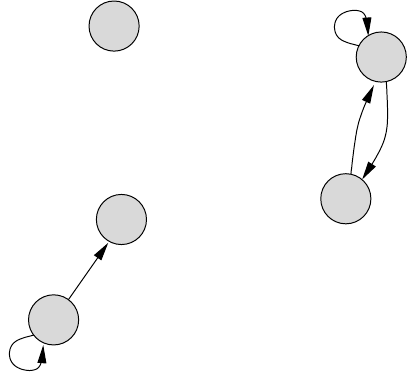}} \\[-40mm]
& $\begin{cases}
\widetilde{f}_1:x\mapsto  0\\
\widetilde{f}_2:x\mapsto  \neg x_3\\
\widetilde{f}_3:x\mapsto  \neg x_3\\
~\\
\widetilde{f}_{5}:x\mapsto   x_{6}\\
\widetilde{f}_{6}:x\mapsto x_{6}\vee x_{5}
\end{cases}$\\[20mm]\end{tabular}
\caption{The BAN $\widetilde{\N}$. {\sc left:}  $\widetilde{\N}$'s interaction graph
$\widetilde{G}=(\widetilde{V},\widetilde{A})$. {\sc right:} $\widetilde{\N}$'s
local update functions.}
\label{fig:BANc}
\end{figure}

% °°°°°°°°°°°°°°°°°°°°°°°°°°°°°°°°°°°°°°°°°°°°°°°°°°°°°°°°°°°°°°°°°°°°°°°°°°°°°°°
% °                                                                             °
% °°°°°°°°°°°°°°°°°°°°°°°°°°°°°°°°°°°°°°°°°°°°°°°°°°°°°°°°°°°°°°°°°°°°°°°°°°°°°°°
To illustrate observations 2 and 3, consider the system represented by BAN
$\widetilde{\N}$ of \figref{fig:BANc}. 

This system is actually three independent systems that we have no reasons to
consider as a whole. Automaton \scalebox{0.4}{\begin{picture}(0,0)%
\includegraphics{automaton1.pdf}%
\end{picture}%
\setlength{\unitlength}{4144sp}%
\begingroup\makeatletter\ifx\SetFigFont\undefined%
\gdef\SetFigFont#1#2#3#4#5{%
  \reset@font\fontsize{#1}{#2pt}%
  \fontfamily{#3}\fontseries{#4}\fontshape{#5}%
  \selectfont}%
\fi\endgroup%
\begin{picture}(398,398)(2097,-935)
\put(2206,-826){\makebox(0,0)[lb]{\smash{{\SetFigFont{20}{24.0}{\familydefault}{\mddefault}{\updefault}{\color[rgb]{0,0,0}1}%
}}}}
\end{picture}%
}{} in particular, is stuck in state
$0$. There is no reason for us to consider \scalebox{0.4}{\begin{picture}(0,0)%
\includegraphics{automaton1.pdf}%
\end{picture}%
\setlength{\unitlength}{4144sp}%
\begingroup\makeatletter\ifx\SetFigFont\undefined%
\gdef\SetFigFont#1#2#3#4#5{%
  \reset@font\fontsize{#1}{#2pt}%
  \fontfamily{#3}\fontseries{#4}\fontshape{#5}%
  \selectfont}%
\fi\endgroup%
\begin{picture}(398,398)(2097,-935)
\put(2206,-826){\makebox(0,0)[lb]{\smash{{\SetFigFont{20}{24.0}{\familydefault}{\mddefault}{\updefault}{\color[rgb]{0,0,0}1}%
}}}}
\end{picture}%
}{} as an {\em interacting}
entity interacting with other entities. Yet in some circumstances, this BAN too
is the perfect representation of what is given of system $\N$ (of
\figref{fig:BANa}) for us to understand.\bigskip

Imagine that entities \scalebox{0.4}{\begin{picture}(0,0)%
\includegraphics{automaton1.pdf}%
\end{picture}%
\setlength{\unitlength}{4144sp}%
\begingroup\makeatletter\ifx\SetFigFont\undefined%
\gdef\SetFigFont#1#2#3#4#5{%
  \reset@font\fontsize{#1}{#2pt}%
  \fontfamily{#3}\fontseries{#4}\fontshape{#5}%
  \selectfont}%
\fi\endgroup%
\begin{picture}(398,398)(2097,-935)
\put(2206,-826){\makebox(0,0)[lb]{\smash{{\SetFigFont{20}{24.0}{\familydefault}{\mddefault}{\updefault}{\color[rgb]{0,0,0}1}%
}}}}
\end{picture}%
}{}, \scalebox{0.4}{\begin{picture}(0,0)%
\includegraphics{automaton2.pdf}%
\end{picture}%
\setlength{\unitlength}{4144sp}%
\begingroup\makeatletter\ifx\SetFigFont\undefined%
\gdef\SetFigFont#1#2#3#4#5{%
  \reset@font\fontsize{#1}{#2pt}%
  \fontfamily{#3}\fontseries{#4}\fontshape{#5}%
  \selectfont}%
\fi\endgroup%
\begin{picture}(398,398)(2097,-935)
\put(2206,-826){\makebox(0,0)[lb]{\smash{{\SetFigFont{20}{24.0}{\familydefault}{\mddefault}{\updefault}{\color[rgb]{0,0,0}2}%
}}}}
\end{picture}%
}{}, \scalebox{0.4}{\begin{picture}(0,0)%
\includegraphics{automaton3.pdf}%
\end{picture}%
\setlength{\unitlength}{4144sp}%
\begingroup\makeatletter\ifx\SetFigFont\undefined%
\gdef\SetFigFont#1#2#3#4#5{%
  \reset@font\fontsize{#1}{#2pt}%
  \fontfamily{#3}\fontseries{#4}\fontshape{#5}%
  \selectfont}%
\fi\endgroup%
\begin{picture}(398,398)(2097,-935)
\put(2206,-826){\makebox(0,0)[lb]{\smash{{\SetFigFont{20}{24.0}{\familydefault}{\mddefault}{\updefault}{\color[rgb]{0,0,0}3}%
}}}}
\end{picture}%
}{}, \scalebox{0.4}{\begin{picture}(0,0)%
\includegraphics{automaton4.pdf}%
\end{picture}%
\setlength{\unitlength}{4144sp}%
\begingroup\makeatletter\ifx\SetFigFont\undefined%
\gdef\SetFigFont#1#2#3#4#5{%
  \reset@font\fontsize{#1}{#2pt}%
  \fontfamily{#3}\fontseries{#4}\fontshape{#5}%
  \selectfont}%
\fi\endgroup%
\begin{picture}(398,398)(2097,-935)
\put(2206,-826){\makebox(0,0)[lb]{\smash{{\SetFigFont{20}{24.0}{\familydefault}{\mddefault}{\updefault}{\color[rgb]{0,0,0}4}%
}}}}
\end{picture}%
}{} of $\N$ happen to be
caught in the same rhythm, and for the same sort of reasons as before, we are unaware of
entity \scalebox{0.4}{\begin{picture}(0,0)%
\includegraphics{automaton4.pdf}%
\end{picture}%
\setlength{\unitlength}{4144sp}%
\begingroup\makeatletter\ifx\SetFigFont\undefined%
\gdef\SetFigFont#1#2#3#4#5{%
  \reset@font\fontsize{#1}{#2pt}%
  \fontfamily{#3}\fontseries{#4}\fontshape{#5}%
  \selectfont}%
\fi\endgroup%
\begin{picture}(398,398)(2097,-935)
\put(2206,-826){\makebox(0,0)[lb]{\smash{{\SetFigFont{20}{24.0}{\familydefault}{\mddefault}{\updefault}{\color[rgb]{0,0,0}4}%
}}}}
\end{picture}%
}{} (\eg we are looking at interactions between varying
concentrations \scalebox{0.4}{\begin{picture}(0,0)%
\includegraphics{automatoni.pdf}%
\end{picture}%
\setlength{\unitlength}{4144sp}%
\begingroup\makeatletter\ifx\SetFigFont\undefined%
\gdef\SetFigFont#1#2#3#4#5{%
  \reset@font\fontsize{#1}{#2pt}%
  \fontfamily{#3}\fontseries{#4}\fontshape{#5}%
  \selectfont}%
\fi\endgroup%
\begin{picture}(398,398)(2097,-935)
\put(2206,-826){\makebox(0,0)[lb]{\smash{{\SetFigFont{20}{24.0}{\familydefault}{\mddefault}{\updefault}{\color[rgb]{0,0,0}$i$}%
}}}}
\end{picture}%
}{} $i\neq 4$ in the cell of different proteins, and are
thereby unable to perceive changes affecting genes like \scalebox{0.4}{\begin{picture}(0,0)%
\includegraphics{automaton4.pdf}%
\end{picture}%
\setlength{\unitlength}{4144sp}%
\begingroup\makeatletter\ifx\SetFigFont\undefined%
\gdef\SetFigFont#1#2#3#4#5{%
  \reset@font\fontsize{#1}{#2pt}%
  \fontfamily{#3}\fontseries{#4}\fontshape{#5}%
  \selectfont}%
\fi\endgroup%
\begin{picture}(398,398)(2097,-935)
\put(2206,-826){\makebox(0,0)[lb]{\smash{{\SetFigFont{20}{24.0}{\familydefault}{\mddefault}{\updefault}{\color[rgb]{0,0,0}4}%
}}}}
\end{picture}%
}\,). 
\medskip

In $\N$, everything is as if updates were being made in the following periodic
order: 

$$\ldots\, 3,2,{\color{black!50}4,}1,3,2,{\color{black!50}4,}1,3,2,{\color{black!50}4,}1\ldots{}\,$$

Imagine also, that we are observing the system $\N$ with regularity. But still,
we are not there absolutely each and every time something in $\N$ changes. More
precisely, imagine that in $\N$ the changes of states of entities happen very
fast in comparison to the whole duration of a period of updates
$(3,2,{\color{black!50}4,}1)$. Because of the that and because of the regularity
of our observations with respect to that, while $\N$ takes trajectory:

\begin{gather*}
\raisebox{1.8pt}{\ldots}\hspace{-4pt}\longrightarrow 
x=(x_1,x_2,x_3,x_4,x_5,x_6)\stackrel{{\color{black!20}\scriptsize 3}}{\longrightarrow}
x'=(x_1,x_2,f_3(x),x_4,x_5,x_6)\stackrel{{\color{black!20}\scriptsize 2}}{\longrightarrow}\hspace{4cm}\\
x''=(x_1,f_2(x'),f_3(x),x_4,x_5,x_6)\stackrel{{\color{black!20}\scriptsize 4}}{\longrightarrow}
x'''=(x_1,f_2(x'),f_3(x),f_4(x''),x_5,x_6)\\\hspace{7cm}\stackrel{{\color{black!20}\scriptsize 1}}{\longrightarrow}
(f_1(x'''),f_2(x'),f_3(x),f_4(x''),x_5,x_6)
\longrightarrow\hspace{-3pt}\raisebox{1.8pt}{\ldots}
\end{gather*}

we observers of reality just see:

\begin{gather*}
\raisebox{1.8pt}{\ldots}\hspace{-4pt}\longrightarrow 
x=(x_1,x_2,x_3,x_5,x_6)\stackrel{{\color{black!20}\scriptsize
1,2,3}}{\longrightarrow}
(f_1(x'''),f_2(x'),f_3(x),x_5,x_6)
\longrightarrow\hspace{-3pt}\raisebox{1.8pt}{\ldots}
%% \raisebox{1.8pt}{\ldots}\hspace{-4pt}\longrightarrow
%% (x_1,x_2,x_3,x_5,x_6)\stackrel{{\color{black!20}\scriptsize 1}}{\longrightarrow}
%% (f_1(x'),x_2,x_3,x_5,x_6)\longrightarrow\hspace{-3pt}\raisebox{1.8pt}{\ldots}.
\end{gather*}

where $\forall x\in\Bn$:
\begin{align*}
f_1(x''')&=f_4(x'')\wedge (\neg f_2(x') \vee \neg
f_3(x))\\&=(f_2(x') \vee f_3(x))\wedge (\neg f_2(x') \vee \neg
f_3(x)) \\&=(f_3(x) \vee f_3(x))\wedge (\neg f_3(x) \vee \neg
f_3(x)) \\ &= 0.
\end{align*}
\bigskip

So under these circumstances, $\widetilde{\N}$ accurately represents all the
information we get out of our absolutely flawless observation of $\N$ {\em under
  those circumstances}. And according to this accurate representation of $\N$,
in particular there exists no interaction between entities \scalebox{0.4}{\begin{picture}(0,0)%
\includegraphics{automaton1.pdf}%
\end{picture}%
\setlength{\unitlength}{4144sp}%
\begingroup\makeatletter\ifx\SetFigFont\undefined%
\gdef\SetFigFont#1#2#3#4#5{%
  \reset@font\fontsize{#1}{#2pt}%
  \fontfamily{#3}\fontseries{#4}\fontshape{#5}%
  \selectfont}%
\fi\endgroup%
\begin{picture}(398,398)(2097,-935)
\put(2206,-826){\makebox(0,0)[lb]{\smash{{\SetFigFont{20}{24.0}{\familydefault}{\mddefault}{\updefault}{\color[rgb]{0,0,0}1}%
}}}}
\end{picture}%
}{} and
\scalebox{0.4}{\begin{picture}(0,0)%
\includegraphics{automaton5.pdf}%
\end{picture}%
\setlength{\unitlength}{4144sp}%
\begingroup\makeatletter\ifx\SetFigFont\undefined%
\gdef\SetFigFont#1#2#3#4#5{%
  \reset@font\fontsize{#1}{#2pt}%
  \fontfamily{#3}\fontseries{#4}\fontshape{#5}%
  \selectfont}%
\fi\endgroup%
\begin{picture}(398,398)(2097,-935)
\put(2206,-826){\makebox(0,0)[lb]{\smash{{\SetFigFont{20}{24.0}{\familydefault}{\mddefault}{\updefault}{\color[rgb]{0,0,0}5}%
}}}}
\end{picture}%
}\,: $(1,5)\notin \widetilde{A}$.\bigskip

Traditionally in the Bioinformatics
literature \citeal{BenAmor2013,uncertainty2,uncertainty3,Streck2012},
at best only three cases are considered for any two entities \scalebox{0.4}{\begin{picture}(0,0)%
\includegraphics{automatoni.pdf}%
\end{picture}%
\setlength{\unitlength}{4144sp}%
\begingroup\makeatletter\ifx\SetFigFont\undefined%
\gdef\SetFigFont#1#2#3#4#5{%
  \reset@font\fontsize{#1}{#2pt}%
  \fontfamily{#3}\fontseries{#4}\fontshape{#5}%
  \selectfont}%
\fi\endgroup%
\begin{picture}(398,398)(2097,-935)
\put(2206,-826){\makebox(0,0)[lb]{\smash{{\SetFigFont{20}{24.0}{\familydefault}{\mddefault}{\updefault}{\color[rgb]{0,0,0}$i$}%
}}}}
\end{picture}%
}{} and \scalebox{0.4}{\begin{picture}(0,0)%
\includegraphics{automatonj.pdf}%
\end{picture}%
\setlength{\unitlength}{4144sp}%
\begingroup\makeatletter\ifx\SetFigFont\undefined%
\gdef\SetFigFont#1#2#3#4#5{%
  \reset@font\fontsize{#1}{#2pt}%
  \fontfamily{#3}\fontseries{#4}\fontshape{#5}%
  \selectfont}%
\fi\endgroup%
\begin{picture}(398,398)(2097,-935)
\put(2206,-826){\makebox(0,0)[lb]{\smash{{\SetFigFont{20}{24.0}{\familydefault}{\mddefault}{\updefault}{\color[rgb]{0,0,0}$j$}%
}}}}
\end{picture}%
}{} of
a  real system $\N$:

\begin{enumerate}\renewcommand{\labelenumi}{\litemunlike{\sc Case\,\arabic{enumi}:}} 

\item Entity \scalebox{0.4}{\begin{picture}(0,0)%
\includegraphics{automatoni.pdf}%
\end{picture}%
\setlength{\unitlength}{4144sp}%
\begingroup\makeatletter\ifx\SetFigFont\undefined%
\gdef\SetFigFont#1#2#3#4#5{%
  \reset@font\fontsize{#1}{#2pt}%
  \fontfamily{#3}\fontseries{#4}\fontshape{#5}%
  \selectfont}%
\fi\endgroup%
\begin{picture}(398,398)(2097,-935)
\put(2206,-826){\makebox(0,0)[lb]{\smash{{\SetFigFont{20}{24.0}{\familydefault}{\mddefault}{\updefault}{\color[rgb]{0,0,0}$i$}%
}}}}
\end{picture}%
}{} {\em really}
impacts on entity \scalebox{0.4}{\begin{picture}(0,0)%
\includegraphics{automatonj.pdf}%
\end{picture}%
\setlength{\unitlength}{4144sp}%
\begingroup\makeatletter\ifx\SetFigFont\undefined%
\gdef\SetFigFont#1#2#3#4#5{%
  \reset@font\fontsize{#1}{#2pt}%
  \fontfamily{#3}\fontseries{#4}\fontshape{#5}%
  \selectfont}%
\fi\endgroup%
\begin{picture}(398,398)(2097,-935)
\put(2206,-826){\makebox(0,0)[lb]{\smash{{\SetFigFont{20}{24.0}{\familydefault}{\mddefault}{\updefault}{\color[rgb]{0,0,0}$j$}%
}}}}
\end{picture}%
}{}, possibly indirectly, and the model $\widetilde{\N}$ of
$\N$ formalises this through the arc $(i,j)\in\widetilde{A}$.

\item Entity \scalebox{0.4}{\begin{picture}(0,0)%
\includegraphics{automatoni.pdf}%
\end{picture}%
\setlength{\unitlength}{4144sp}%
\begingroup\makeatletter\ifx\SetFigFont\undefined%
\gdef\SetFigFont#1#2#3#4#5{%
  \reset@font\fontsize{#1}{#2pt}%
  \fontfamily{#3}\fontseries{#4}\fontshape{#5}%
  \selectfont}%
\fi\endgroup%
\begin{picture}(398,398)(2097,-935)
\put(2206,-826){\makebox(0,0)[lb]{\smash{{\SetFigFont{20}{24.0}{\familydefault}{\mddefault}{\updefault}{\color[rgb]{0,0,0}$i$}%
}}}}
\end{picture}%
}{} {\em
really} has no influence on entity \scalebox{0.4}{\begin{picture}(0,0)%
\includegraphics{automatonj.pdf}%
\end{picture}%
\setlength{\unitlength}{4144sp}%
\begingroup\makeatletter\ifx\SetFigFont\undefined%
\gdef\SetFigFont#1#2#3#4#5{%
  \reset@font\fontsize{#1}{#2pt}%
  \fontfamily{#3}\fontseries{#4}\fontshape{#5}%
  \selectfont}%
\fi\endgroup%
\begin{picture}(398,398)(2097,-935)
\put(2206,-826){\makebox(0,0)[lb]{\smash{{\SetFigFont{20}{24.0}{\familydefault}{\mddefault}{\updefault}{\color[rgb]{0,0,0}$j$}%
}}}}
\end{picture}%
}{} and the model accounts for this through the
absence of arc $(i,j)\notin \widetilde{A}$. 

\item \label{case3} Entity \scalebox{0.4}{\begin{picture}(0,0)%
\includegraphics{automatoni.pdf}%
\end{picture}%
\setlength{\unitlength}{4144sp}%
\begingroup\makeatletter\ifx\SetFigFont\undefined%
\gdef\SetFigFont#1#2#3#4#5{%
  \reset@font\fontsize{#1}{#2pt}%
  \fontfamily{#3}\fontseries{#4}\fontshape{#5}%
  \selectfont}%
\fi\endgroup%
\begin{picture}(398,398)(2097,-935)
\put(2206,-826){\makebox(0,0)[lb]{\smash{{\SetFigFont{20}{24.0}{\familydefault}{\mddefault}{\updefault}{\color[rgb]{0,0,0}$i$}%
}}}}
\end{picture}%
}{} {\em really} impacts on entity \scalebox{0.4}{\begin{picture}(0,0)%
\includegraphics{automatonj.pdf}%
\end{picture}%
\setlength{\unitlength}{4144sp}%
\begingroup\makeatletter\ifx\SetFigFont\undefined%
\gdef\SetFigFont#1#2#3#4#5{%
  \reset@font\fontsize{#1}{#2pt}%
  \fontfamily{#3}\fontseries{#4}\fontshape{#5}%
  \selectfont}%
\fi\endgroup%
\begin{picture}(398,398)(2097,-935)
\put(2206,-826){\makebox(0,0)[lb]{\smash{{\SetFigFont{20}{24.0}{\familydefault}{\mddefault}{\updefault}{\color[rgb]{0,0,0}$j$}%
}}}}
\end{picture}%
}{} but the
experimental data collected upstream by biologists has failed to evidence this
fact about reality. As a consequence,  the theory is failing to represent it: arc
$(i,j)\in\widetilde{A}$ is accidentally missing from $\widetilde{\N}$'s
interaction graph $\widetilde{G}=(\widetilde{V},\widetilde{A})$.
\end{enumerate}

In agreement with Observation 2, the example of BAN $\widetilde{\N}$
of \figref{fig:BANc} modelling system $\N$ of \figref{fig:BANa} shows
that \litemunlike{{\sc Case}\,2} doesn't make any sense at all beyond the
`reality' of a specific level of abstraction (\eg the one at which
concentrations \scalebox{0.4}{\begin{picture}(0,0)%
\includegraphics{automatoni.pdf}%
\end{picture}%
\setlength{\unitlength}{4144sp}%
\begingroup\makeatletter\ifx\SetFigFont\undefined%
\gdef\SetFigFont#1#2#3#4#5{%
  \reset@font\fontsize{#1}{#2pt}%
  \fontfamily{#3}\fontseries{#4}\fontshape{#5}%
  \selectfont}%
\fi\endgroup%
\begin{picture}(398,398)(2097,-935)
\put(2206,-826){\makebox(0,0)[lb]{\smash{{\SetFigFont{20}{24.0}{\familydefault}{\mddefault}{\updefault}{\color[rgb]{0,0,0}$i$}%
}}}}
\end{picture}%
}{} $i\neq 4$ of proteins in the cell are meaningful and
visible, and a
gene  \scalebox{0.4}{\begin{picture}(0,0)%
\includegraphics{automaton4.pdf}%
\end{picture}%
\setlength{\unitlength}{4144sp}%
\begingroup\makeatletter\ifx\SetFigFont\undefined%
\gdef\SetFigFont#1#2#3#4#5{%
  \reset@font\fontsize{#1}{#2pt}%
  \fontfamily{#3}\fontseries{#4}\fontshape{#5}%
  \selectfont}%
\fi\endgroup%
\begin{picture}(398,398)(2097,-935)
\put(2206,-826){\makebox(0,0)[lb]{\smash{{\SetFigFont{20}{24.0}{\familydefault}{\mddefault}{\updefault}{\color[rgb]{0,0,0}4}%
}}}}
\end{picture}%
}{} isn't).\bigskip

In agreement with Observation 3, the example also evidences there can be other
reasons --~different from the ``{unfortunate data deficit}''
underpinning \litemunlike{{\sc Case}\,3}~-- for the representation
$\widetilde{\N}$ of a system $\N$ to fail to account for interactions between
entities. \bigskip 

Now, consider again $\widetilde{\N}$ -- the BAN of \figref{fig:BANc}
representing precisely what we see of the system $\N$  of \figref{fig:BANa} under the conditions described above.
Under the conditions described above, if one entity of $\N$ were to change pace,
slightly slow down for instance, even if only momentarily, then \scalebox{0.4}{\begin{picture}(0,0)%
\includegraphics{automaton1.pdf}%
\end{picture}%
\setlength{\unitlength}{4144sp}%
\begingroup\makeatletter\ifx\SetFigFont\undefined%
\gdef\SetFigFont#1#2#3#4#5{%
  \reset@font\fontsize{#1}{#2pt}%
  \fontfamily{#3}\fontseries{#4}\fontshape{#5}%
  \selectfont}%
\fi\endgroup%
\begin{picture}(398,398)(2097,-935)
\put(2206,-826){\makebox(0,0)[lb]{\smash{{\SetFigFont{20}{24.0}{\familydefault}{\mddefault}{\updefault}{\color[rgb]{0,0,0}1}%
}}}}
\end{picture}%
}{} might at
some point take state $1$. The oscillations of \scalebox{0.4}{\begin{picture}(0,0)%
\includegraphics{automaton3.pdf}%
\end{picture}%
\setlength{\unitlength}{4144sp}%
\begingroup\makeatletter\ifx\SetFigFont\undefined%
\gdef\SetFigFont#1#2#3#4#5{%
  \reset@font\fontsize{#1}{#2pt}%
  \fontfamily{#3}\fontseries{#4}\fontshape{#5}%
  \selectfont}%
\fi\endgroup%
\begin{picture}(398,398)(2097,-935)
\put(2206,-826){\makebox(0,0)[lb]{\smash{{\SetFigFont{20}{24.0}{\familydefault}{\mddefault}{\updefault}{\color[rgb]{0,0,0}3}%
}}}}
\end{picture}%
}{} might even spread to
\scalebox{0.4}{\begin{picture}(0,0)%
\includegraphics{automaton5.pdf}%
\end{picture}%
\setlength{\unitlength}{4144sp}%
\begingroup\makeatletter\ifx\SetFigFont\undefined%
\gdef\SetFigFont#1#2#3#4#5{%
  \reset@font\fontsize{#1}{#2pt}%
  \fontfamily{#3}\fontseries{#4}\fontshape{#5}%
  \selectfont}%
\fi\endgroup%
\begin{picture}(398,398)(2097,-935)
\put(2206,-826){\makebox(0,0)[lb]{\smash{{\SetFigFont{20}{24.0}{\familydefault}{\mddefault}{\updefault}{\color[rgb]{0,0,0}5}%
}}}}
\end{picture}%
}{}. Or, if \scalebox{0.4}{\begin{picture}(0,0)%
\includegraphics{automaton5.pdf}%
\end{picture}%
\setlength{\unitlength}{4144sp}%
\begingroup\makeatletter\ifx\SetFigFont\undefined%
\gdef\SetFigFont#1#2#3#4#5{%
  \reset@font\fontsize{#1}{#2pt}%
  \fontfamily{#3}\fontseries{#4}\fontshape{#5}%
  \selectfont}%
\fi\endgroup%
\begin{picture}(398,398)(2097,-935)
\put(2206,-826){\makebox(0,0)[lb]{\smash{{\SetFigFont{20}{24.0}{\familydefault}{\mddefault}{\updefault}{\color[rgb]{0,0,0}5}%
}}}}
\end{picture}%
}{} and \scalebox{0.4}{\begin{picture}(0,0)%
\includegraphics{automaton6.pdf}%
\end{picture}%
\setlength{\unitlength}{4144sp}%
\begingroup\makeatletter\ifx\SetFigFont\undefined%
\gdef\SetFigFont#1#2#3#4#5{%
  \reset@font\fontsize{#1}{#2pt}%
  \fontfamily{#3}\fontseries{#4}\fontshape{#5}%
  \selectfont}%
\fi\endgroup%
\begin{picture}(398,398)(2097,-935)
\put(2206,-826){\makebox(0,0)[lb]{\smash{{\SetFigFont{20}{24.0}{\familydefault}{\mddefault}{\updefault}{\color[rgb]{0,0,0}6}%
}}}}
\end{picture}%
}{} had been locked in state $0$ until then,
\scalebox{0.4}{\begin{picture}(0,0)%
\includegraphics{automaton1.pdf}%
\end{picture}%
\setlength{\unitlength}{4144sp}%
\begingroup\makeatletter\ifx\SetFigFont\undefined%
\gdef\SetFigFont#1#2#3#4#5{%
  \reset@font\fontsize{#1}{#2pt}%
  \fontfamily{#3}\fontseries{#4}\fontshape{#5}%
  \selectfont}%
\fi\endgroup%
\begin{picture}(398,398)(2097,-935)
\put(2206,-826){\makebox(0,0)[lb]{\smash{{\SetFigFont{20}{24.0}{\familydefault}{\mddefault}{\updefault}{\color[rgb]{0,0,0}1}%
}}}}
\end{picture}%
}{} might unlock \scalebox{0.4}{\begin{picture}(0,0)%
\includegraphics{automaton5.pdf}%
\end{picture}%
\setlength{\unitlength}{4144sp}%
\begingroup\makeatletter\ifx\SetFigFont\undefined%
\gdef\SetFigFont#1#2#3#4#5{%
  \reset@font\fontsize{#1}{#2pt}%
  \fontfamily{#3}\fontseries{#4}\fontshape{#5}%
  \selectfont}%
\fi\endgroup%
\begin{picture}(398,398)(2097,-935)
\put(2206,-826){\makebox(0,0)[lb]{\smash{{\SetFigFont{20}{24.0}{\familydefault}{\mddefault}{\updefault}{\color[rgb]{0,0,0}5}%
}}}}
\end{picture}%
}{} which in turn might generate the irrevocable
effect of allowing \scalebox{0.4}{\begin{picture}(0,0)%
\includegraphics{automaton6.pdf}%
\end{picture}%
\setlength{\unitlength}{4144sp}%
\begingroup\makeatletter\ifx\SetFigFont\undefined%
\gdef\SetFigFont#1#2#3#4#5{%
  \reset@font\fontsize{#1}{#2pt}%
  \fontfamily{#3}\fontseries{#4}\fontshape{#5}%
  \selectfont}%
\fi\endgroup%
\begin{picture}(398,398)(2097,-935)
\put(2206,-826){\makebox(0,0)[lb]{\smash{{\SetFigFont{20}{24.0}{\familydefault}{\mddefault}{\updefault}{\color[rgb]{0,0,0}6}%
}}}}
\end{picture}%
}{} to take state $1$. With our perspective on $\N$, none
of this would fit with what we know. Worse, we would be essentially unable to foresee
and even understand any of it if it happened.\bigskip\bigskip\bigskip\bigskip

% °°°°°°°°°°°°°°°°°°°°°°°°°°°°°°°°°°°°°°°°°°°°°°°°°°°°°°°°°°°°°°°°°°°°°°°°°°°°°°°
% °                                                                             °
% °°°°°°°°°°°°°°°°°°°°°°°°°°°°°°°°°°°°°°°°°°°°°°°°°°°°°°°°°°°°°°°°°°°°°°°°°°°°°°°
\begin{observationG}
What information a formal object can provide about the real system it is meant
   to model, and what uninterpretable, non-modelling information it provides on
 top of that, does not just depend on the semantics associated to the formalism
 describing the formal object,
 it also essentially depends on the relative consistency of the semantics.
\label{obs:MTG}
\end{observationG}

\begin{observationG}
The semantics associated to a formal object can be decisively affected by the
history of the object and how our scientific community came to inherit it. And
thus, so can the precise definition of the object that we choose among different
customary variations of the definition, and the properties it has that we
take interest in, and those we don't. \linebreak\vspace{-4mm}
\label{obs:comG}
\end{observationG}

% °°°°°°°°°°°°°°°°°°°°°°°°°°°°°°°°°°°°°°°°°°°°°°°°°°°°°°°°°°°°°°°°°°°°°°°°°°°°°°°
% °                                                                             °
% °°°°°°°°°°°°°°°°°°°°°°°°°°°°°°°°°°°°°°°°°°°°°°°°°°°°°°°°°°°°°°°°°°°°°°°°°°°°°°°

The last two observations can be illustrated with the notion of
``synchronism''.\bigskip

A surprisingly great many occidental modellers of biological regulation networks
confuse synchronism in BANs with the parallel update schedule (PUS) of BANs
\citeal{Bernot2004,Vijesh2013,ConfuseParallelAndSynchronism2,ConfuseParallelAndSynchronism3,ConfuseParallelAndSynchronism4,ConfuseParallelAndSynchronism7Nice,ConfuseParallelAndSynchronism5Naldi,ConfuseParallelAndSynchronism6Naldi,Delaplace2010,HDeJongReview,ConfuseParallelAndSynchronism10,ConfuseParallelAndSynchronism11,ConfuseParallelAndSynchronism13Naldi,ConfuseParallelAndSynchronism12,Thomas1990B}.
The PUS is the update schedule originally used and made sense of by McCulloch
and Pitts in their seminal BANs \citeal{McCulloch1943}. 
The PUS forces a BAN to
systematically update all its automata so that $\forall x\in \Bn$, the BAN
transitions from $x$ to $(f_1(x),\ldots,f_n(x))\in \Bn$.
When this makes some automata react more quickly than we would like them to,
intermediary automata can simply be added as it was originally done in the
McCulloch and Pitts BANs.
Asynchronism, to which the PUS is wrongly opposed, is the update constraint
that rules out the possibility of updating {more than one} automaton in $x$.
Non-asynchronism, a.k.a synchronism is the possibility of updating more than one
in $x$. Formally, it is expressed by: $|U(x)|>1$ (more than $1$ automata can
change states in $x$).\linebreak
\bigskip

Such great tenacity for such a coarse confusion can only be explained by the 
fact that wherever it is made, {\em it does not matter}, or at least, {\em
it is not made to matter}.\linebreak
A widespread confusion is nonetheless still is a confusion. Since it is widespread, in
agreement with Observation 5, it is much more likely to be the legacy of a community
blind spot induced by inherited semantics, than to
be the responsibility of individual err.\linebreak
And indeed, the blind
spot around synchronism seems to be a natural effect of the constant diverse
historical reprocessing of BANs and of the way sense is made out of
them \citeal{McCulloch1943,Goles1990B,Hopfield1982,IsingNeural,Kauffman1969,Thomas1973}.
Extensive interdisciplinary interest in BANs has been reassigning modelling
responsibilities to BANs faster then it has been answering questions about what
BANs {\em can}, and actually {\em do}  formalise.\bigskip
% on associe des pouvoirs de modellisations plus vote qu'on se pose la question de
% ce que les BANs sont capables de formaliser et ce 

\begin{figure}[h!]
\centerline{\fcolorbox{gray}{black!5}{\scalebox{0.6}{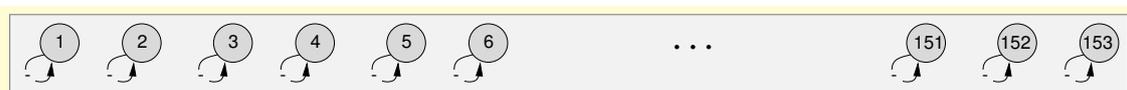}}}
\caption{The interaction graph $G^{\ast}$ of BAN $\N^\ast=\set{f_i:x\mapsto \neg x_i,\, \forall i\in \llbracket
  1,153\rrbracket}$.}
\label{fig:BAN153}
\end{figure}

The confusion  between the parallel update schedule and the notion of
synchronism results in: 
\begin{enumerate}\renewcommand{\labelenumi}{\litemlike{\arabic{enumi}}}
\item The neglect of all intermediary updating possibilities
that neither rule out synchronism altogether, nor rule out asynchronism
altogether (around $2^{153+2^{153}}$ in the case of BAN $\N^{\ast}$
of \figref{fig:BAN153}), ~and
\item \textls[-5]{Research being confined to frameworks in which synchronism is never
  considered independently of the other very strong characteristic features of
  the PUS (\eg determinism, periodicity) for the reason that in those
  frameworks, it {\em cannot} be.}\linebreak
\end{enumerate}

\bigskip

The disregard synchronism owes in particular to its misguided association with
determinism, is aggravated by two assumptions commonly used to motivate an
asynchronous updating in some communities that are interested in modelling genetic regulation with
(B)ANs \citeal{Thomas1991}:
%\linebreak\vspace{-4mm}

\itshape
\begin{enumerate}\renewcommand{\labelenumi}{\litemlike{\arabic{enumi}}}
\item Simultaneity in nature is highly fortuitous, ~\normalfont and \itshape
\item Simultaneity in nature  maps bijectively onto synchronism in
(B)ANs.
\end{enumerate}
\normalfont

The notion of  {\em simultaneity} implied in these assumptions requires a notion of
``{objective time}'' to make sense.  {\it A priori}, in BANs, it doesn't. At least
not spontaneously.  What synchronism in BANs conveys is the {absence} of a
causal relation: \medskip

\centerline{\em Synchronously possible events are events that don't need one another
to occur.  }\medskip

So the relation of synchronism relates possible events without specifying
anything about how these events are otherwise related -- that is, how they are
related otherwise than by the relation created by their synchronous
possibility. In particular, {knowing only that events A and B can occur
  synchronously}, means: {not knowing what effect the occurrence of A might have
  on B and {\it vice versa}}\footnote{This absence of information conveyed by
  the relation of synchronism is a typical example of absence of information
  traditionally getting outshined by specialised knowledge and the assumptions
  inspired by specialised knowledge. 
 If anything, what this absence of information represents is ''wriggle room''
 (see \myciteOL{Caus} {Section 9}). And making sense out of it calls for the
 careful attention of Computer Science with its fundamental ability to soundly
 manipulate representations of information and systems, much more than it is
 calls for the other natural sciences' specialised detailed knowledge about the
 complexity and diversity of real life systems in need of modelling.}.
Forcing asynchronism in BANs -- \ie forcing asynchronism in $x$, $\forall x\in
\Bn$ despite $|U(x)|>1$ possibly being true in some $x$ -- contradicts the fact
mentioned above (``{\em In BANs synchronously possible events are events that
  don't need one another to occur}'').  To rule out the possibility of the
synchronous occurrences of possible events by imposing asynchronous updatings,
is to assume that the occurrence of any event prevents the occurrence of all
other that are also possible -- {\em or else that the model conveys the
  causality that we expect very poorly}.  In an asynchronous BAN, automata that
find themselves synchronously unstable are therefore necessarily automata that
have the ability of influencing each other. For the BAN $\N^{\ast}$ of
\figref{fig:BAN153}, this means that 23256 arcs are missing from $G^{\ast}$.
This inconsistency between what we want the BAN to mean and what we want its
constituent updating schedule to mean squarely dismantles the intrinsically
discrete modelling capacity of the mathematical formalism of BANs and severely stakes
the relevance of any information we might subsequently draw out of it about a
real system. In agreement with Observation 4, this shows that overloading
formalism with incompatible semantics is generally {\em not} innocuous.

\section{Conclusion}

The observations and examples given above call for a shift of attention from
specifics and realism to definition and consistency. They show the need for us
to systematically endeavour to refine and update our scientific views so that
instead of speaking of {\em theory} and {\em formalism} as opposed to {\em
reality} and {\em nature}, we rather speak of objects that are abstractions of
one another in a sense of the term ``abstraction'' that we can actually formally
explicit.
\bigskip

A characteristic strength (and beauty) of science is its ability and tradition
of tackling problems and questions through many different angles. Science does
not especially aspire at a one-dimensional history of science-making.  Different
perspectives currently upheld by different contemporary scientific communities
may co-exist. They don't need to mutually invalidate one another since science
doesn't especially need a single consistent ``survivor scientific perspective''
to be selected in the end.  So if science-making presently makes sense in
itself, then the same way, having different scientific communities upholding
different scientific perspectives on the same objects must presently make sense
in itself too.  We have yet to study and explicit the
general coherence there must thereby be in the present coexistence of those
perspectives\footnotemark. This paper
suggests that the minimalist formalism of BANs may be of value in that.
\bigskip

\footnotetext{Incidentally, this  differs  significantly from what the
academic trend of interdisciplinarity is actually proposing.  As far as Computer
Science/Biology collaborations around BANs go, interdisciplinary operates as
discipline-concatenation, prejudicially opposing a conglomerate notion of
``fundamental-/theoretical-ness'' to a notion of ``applied-ness''.  From it
ensues an effective separation between specialists' contributions: the work of
one starts where the work of the other ends. The fatalistic view on information
lack captured by \litemunlike{{\sc Case}\,3} on Page \pageref{case3} --
information lack assumed to be necessarily unfortunate and accidental -- is
bound to consolidate under such circumstances: the validity of the models built
by computer scientists for biologists is conditional to the quality and
completeness of data that computer scientists have no say on.  Under such
circumstances, making the interests of distinct academic fields coincide rather
than punctually relay each other, becomes a matter of educating the affiliates
of one discipline to the specialised knowledge of the other, or else inventing a
new discipline circumscribed to  their intersection.  Interdisciplinarity's
proposition therefore draws its sense out of the fact that specialised academic
knowledge reinforces interdisciplinary separation. This paper means to emphasise
issues that draw their meaning before any  differences between
disciplines need be considered.}

Science has been concentrating on a certain kind of information, emerging
directly from objects, from the explicit statements we make about them.
%%   But in order
%% to ensure a healthy and optimal progression of science, the meaning and the
%% precise formalisation of this information need to be tamed.
The BAN examples above suggest that if we want to rely on the representations we
make for ourselves of the objects we take interest in, then we need to have 
practical in-depth understanding of how a piece of information's  meaning and
formalisation relate to one another.\linebreak
The five observations of this paper undercut the assumption that there are
representations that are fundamentally more ``objective'' -- as in more
``accurate'', more ``complete'' or more ``realistic'' -- than others. 
In that,
they undercut the necessity of restricting ourselves to looking for new
information where beacons the information we already have precisely as we
presently represent it and precisely as we presently interpret it.
%
%% There is another kind of information requiering thorough methodical
%% exploration. It is a virtual kind of information overshadowed by what we already
%% know on the objects it can tell us more about. It is 
%% %that is not yet in communicable form, 
%%  {not} accounted for in the ``data deficit'' we tend to be overly aware of. But
%%  it nonetheless is already being tapped into instinctively and extemporaneously
%%  by scientists as evidenced by their chronic and fruitful perspective
%%  recasts.
 \bigskip

Let us assume the following. \litema There is such a thing as changes of
perspective that makes sense with respect to science-making. In other terms,
there are {\em differences} in perspectives that are meaningful in
science. \litema A change of perspectives on object O, from perspective A to
perspective B, does not systematically result from acquiring new information
about O under A and simply {\em increasing} the set of features we take into
account when we consider O. In other terms, the {\em difference} between two
perspectives on the same object is not necessarily information explicitly
expressed about that object.  And if it's not then it raises the following
question about the new information we get on O as a result of seeing O with
perspective B instead of perspective A. Other than waiting for a fortuitous change
of perspective, concentrating on clearing data deficits in the meantime at 
the risk of ending up mislead by a series of questions and answers that only are
relevant under a perspective that isn't \ldots{} {\em How do we access this new
  information on O given that it is not (fully) dependent on what we know of O under A?}
\bigskip

This paper suggests that a first step to answer this question is to tame
implicit information of the kind that explains how BANs $\widehat{\N}$ and
$\widetilde{\N}$ of \figref{fig:BANb} and \figref{fig:BANc} can both be flawless
representations of the BAN $\N$ of \figref{fig:BANa}, and more generally
understand the possibilities and the limits of what one given formal
representation can actually represent.
%

%% the information characterising the difference between a new perspective
%% and an old one is generally not the kind of information we consider as explicit
%% information, emerging directly from what we know about the systems we study, at
%% least not before we change from the old to the new perspective.
%
%% And if it's not, then it is
%% %
%% By understanding the possibilities and the limits of what one given formal
%% representation can actually represent, we can have access to this kind of
%% information, and more importantly, we can have access to the new information
%% about the system that we have thanks to the new perspective. 
%%  To some extent we might be able to uncover deliberately the new information we owe to
%% a new perspective.
\bigskip\bigskip

% °°°°°°°°°°°°°°°°°°°°°°°°°°°°°°°°°°°°°°°°°°°°°°°°°°°°°°°°°°°°°°°°°°°°°°°°°°°°°°°
% °                                                                             °
% °°°°°°°°°°°°°°°°°°°°°°°°°°°°°°°°°°°°°°°°°°°°°°°°°°°°°°°°°°°°°°°°°°°°°°°°°°°°°°°
\renewcommand{\refname}{Bibliography}

\bibliographystyle{unsrt}
{\small \bibliography{bibliofr}}

\begin{thebibliography}{10}

\bibitem{Barrett2001}
C.~L. Barrett, H.~B. Hunt, M.~V. Marathe, S.~S. Ravi, D.~J. Rosenkrantz, R.~E.
  Stearns, and P.~T. Tosic.
\newblock {Gardens of Eden and fixed points in sequential dynamical systems}.
\newblock In {\em DM-CCG}, 2001.

\bibitem{ColonReyes2004}
O.~Col{\'o}n-Reyes, R.~Laubenbacher, and B.~Pareigis.
\newblock {Boolean monomial dynamical systems}.
\newblock {\em Ann. Comb.}, 8:425--439, 2004.

\bibitem{Delaplace2011}
F.~Delaplace, H.~Klaudel, T.~Melliti, and S.~Sen{\'e}.
\newblock \href{http://arxiv.org/abs/1111.2313}{Modular organisation of
  interaction networks based on asymptotic dynamics}.
\newblock arXiv:1111.2313, 2011.

\bibitem{Demongeot2000}
J.~Demongeot, J.~Aracena, Ben~Lamine S., and {\it et al.}
\newblock {\em {Comparative genomics: empirical and analytical approaches to
  gene order dynamics, map alignment and the evolution of gene families}},
  chapter {Hot spots in chromosomal breakage: from description to etiology}.
\newblock Kluwer Academic Publishers, 2000.

\bibitem{Goles1990B}
E.~Goles and S.~Mart{\'i}nez.
\newblock {\em {Neural and automata networks: dynamical behaviour and
  applications}}.
\newblock Kluwer Academic Publishers, 1990.

\bibitem{Goles2008}
E.~Goles{,}~L. Salinas.
\newblock {Comparison between parallel and serial dynamics of Boolean
  networks}.
\newblock {\em Theor. Comp. Sci.}, 396:247--253, 2008.

\bibitem{Jarrah2010}
A.~J. Jarrah, R.~Laubenbacher, and A.~Veliz-Cuba.
\newblock {The dynamics of conjunctive and disjunctive Boolean network models}.
\newblock {\em Bulletin of Mathematical Biology}, 72:1425--1447, 2010.

\bibitem{ConfuseParallelAndSynchronism6Naldi}
D.~Bérenguier{,} C.~Chaouiya {\it et al.}
\newblock Dynamical modeling and analysis of large cellular regulatory
  networks.
\newblock {\em Chaos}, 23(2), 2013.

\bibitem{Mortveit2001}
H.S. Mortveit{,}~C.M. Reidys.
\newblock {Discrete, sequential dynamical systems}.
\newblock {\em Discrete Math.}, 226:281--295, 2001.

\bibitem{MortveitCalculus}
H.S. Mortveit and C.M. Reidys.
\newblock Towards a calculus of biological networks.
\newblock {\em Z. Phys. Chem.}, 216:1--13, 2001.

\bibitem{Aracena2009}
J.~Aracena{,} E. Goles{,} A. Moreira{,}~L. Salinas.
\newblock {On the robustness of update schedules in Boolean networks}.
\newblock {\em Biosystems}, 97:1--8, 2009.

\bibitem{Mendoza1998}
L.~Mendoza and E.~R. Alvarez-Buylla.
\newblock {Dynamics of the genetic regulatory network for Arabidopsis thaliana
  flower morphogenesis}.
\newblock {\em Journal of Theoretical Biology}, 193:307--319, 1998.

\bibitem{Remy2008a}
{\'E}.~Remy, P.~Ruet, and D.~Thieffry.
\newblock {Graphic requirement for multistability and attractive cycles in a
  Boolean dynamical framework}.
\newblock {\em Advances in Applied Mathematics}, 41:335--350, 2008.

\bibitem{Remy2008b}
{\'E}.~Remy{,}~P. Ruet.
\newblock {From minimal signed circuits to the dynamics of Boolean regulatory
  networks}.
\newblock {\em Bioinformatics}, 24:i220--i226, 2008.

\bibitem{Remy2003}
{\'E}.~Remy, B.~Moss{\'e}, C~Chaouiya, and D.~Thieffry.
\newblock {A description of dynamical graphs associated to elementary
  regulatory circuits}.
\newblock {\em Bioinformatics}, 19:ii172--ii178, 2003.

\bibitem{Richard2007}
A.~Richard and J.-P. Comet.
\newblock {Necessary conditions for multistationarity in discrete dynamical
  systems}.
\newblock {\em Discrete Applied Mathematics}, 155:2403--2413, 2007.

\bibitem{Richard2009}
A.~Richard.
\newblock {Positive circuits and maximal number of fixed points in discrete
  dynamical systems}.
\newblock {\em Discrete Applied Mathematics}, 157:3281--3288, 2009.

\bibitem{Robert1986B}
F.~Robert.
\newblock {\em {Discrete iterations: a metric study}}.
\newblock Springer Verlag, 1986.

\bibitem{Siebert2009}
H.~Siebert.
\newblock {Dynamical and structural modularity of discrete regulatory
  networks}.
\newblock In {\em COMPMOD}, volume~6 of {\em Electronic Proceedings in
  Theoretical Computer Science}, 2009.

\bibitem{Tournier2009}
L.~Tournier and M.~Chaves.
\newblock {Uncovering operational interactions in genetic networks using
  asynchronous Boolean dynamics}.
\newblock {\em Journal of Theoretical Biology}, 260:196--209, 2009.

\bibitem{DAM2010}
J.~Demongeot, M.~Noual, and S.~Sen{\'e}.
\newblock
  \mineb\href{http://www.sciencedirect.com/science/article/pii/S0166218X110042%
39}{Combinatorics of Boolean automata circuits dynamics}\minee.
\newblock {\em
  \href{http://www.journals.elsevier.com/discrete-applied-mathematics}{Discrete
  Applied Mathematics}}, 160:398--415, 2010.

\bibitem{AAM2011}
E.~Goles and M.~Noual.
\newblock
  \mineb\href{http://www.sciencedirect.com/science/article/pii/S01968858120001%
39}{Disjunctive networks and update schedules}\minee.
\newblock {\em
  \href{http://www.journals.elsevier.com/advances-in-applied-mathematics}{Adva%
nces in Applied Mathematics}}, 48(5):646 -- 662, 2012.

\bibitem{BMB2013}
J.-P. Comet, M.~Noual, A.~Richard, J.~Aracena, L.~Calzone, D.~Demongeot,
  M.~Kaufman, A.~Naldi, E.H. Snoussi, and D.~Thieffry.
\newblock
  \mineb\href{http://link.springer.com/article/10.1007\%2Fs11538-013-9829-2}{On
  circuit functionality in Boolean networks}\minee.
\newblock {\em \href{http://link.springer.com/journal/11538}{Bulletin of
  Mathematical Biology}}, 75(6):906--19, 2013.

\bibitem{DAM2011}
J.~Aracena, {\'E}.~Fanchon, M.~Montalva, and M.~Noual.
\newblock
  \mineb\href{http://www.sciencedirect.com/science/article/pii/S0166218X100035%
25}{Combinatorics on update digraphs in Boolean networks}\minee.
\newblock {\em
  \href{http://www.journals.elsevier.com/discrete-applied-mathematics}{Discrete
  Applied Mathematics}}, 159:401--409, 2011.

\bibitem{PLOS2010}
J.~Demongeot, E.~Goles, M.~Morvan, M.~Noual, and S.~Sen{\'e}.
\newblock
  \mineb\href{http://dx.doi.org/10.1371\%2Fjournal.pone.0011793}{Attraction
  basins as gauges of robustness against boundary conditions in biological
  complex systems}\minee.
\newblock {\em \href{http://journals.plos.org/plosone/}{PLoS One}}, 5:e11793,
  2010.

\bibitem{IJMS2009}
J.~Demongeot, H.~Ben~Amor, A.~Elena, P.~Gillois, M.~Noual, and S.~Sen{\'e}.
\newblock \mineb\href{http://www.mdpi.com/1422-0067/10/10/4437}{Robustness in
  regulatory interaction networks. A generic approach with applications at
  different levels: physiologic, metabolic and genetic}\minee.
\newblock {\em \href{http://www.mdpi.com/journal/ijms}{International Journal of
  Molecular Sciences}}, 10:4437--4473, 2009.

\bibitem{JTB2011}
J.~Demongeot, A.~Elena, M.~Noual, S.~Sen{\'e}, and F.~Thuderoz.
\newblock
  \mineb\href{http://www.sciencedirect.com/science/article/pii/S00225193110016%
9X}{"Immunetworks", intersecting circuits and dynamics}\minee.
\newblock {\em
  \href{http://www.journals.elsevier.com/journal-of-theoretical-biology}{Journ%
al of Theoretical Biology}}, 280:19--33, 2011.

\bibitem{Demongeot2008c}
J.~Demongeot, M.~Morvan, and S.~Sen{\'e}.
\newblock {Robustness of dynamical systems attraction basins against state
  perturbations: theoretical protocol and application in systems biology}.
\newblock In {\em CISIS}, 2008.

\bibitem{Chaouiya2004}
C.~Chaouiya, E.~Remy, P.~Ruet, and D.~Thieffry.
\newblock {\em ICATPN}, chapter
  \href{http://dx.doi.org/10.1007/978-3-540-27793-4_9}{Qualitative Modelling of
  Genetic Networks: From Logical Regulatory Graphs to Standard Petri Nets},
  pages 137--156.
\newblock Springer Berlin Heidelberg, 2004.

\bibitem{Aracena200649}
J.~Aracena, M.~Gonz{\'a}lez, A.~Zuniga, M.~A. Mendez, and V.~Cambiazo.
\newblock
  \href{http://www.sciencedirect.com/science/article/pii/S0022519305003139}{Re%
gulatory network for cell shape changes during Drosophila ventral furrow
  formation}.
\newblock {\em Journal of Theoretical Biology}, 239(1):49 -- 62, 2006.

\bibitem{Blanchini000562}
F.~Blanchini, E.~Franco, and G.~Giordano.
\newblock \href{http://biorxiv.org/content/early/2013/11/18/000562}{A
  structural classification of candidate oscillators and multistationary
  systems}.
\newblock {\em bioRxiv}, 2013.

\bibitem{GolesOlivos81}
E.~Goles and J.~Olivos.
\newblock \href{http://dx.doi.org/10.1016/S0019-9958(81)90208-4}{The
  Convergence of Symmetric Threshold Automata}.
\newblock {\em Information and Control}, 51(2):98--104, 1981.

\bibitem{Laubenbacher2012B}
R.~Laubenbacher, A.S. Jarrah, H.S. Mortveit, and S.S. Ravi.
\newblock {\em Computational Complexity: Theory, Techniques, and Applications},
  chapter \href{http://dx.doi.org/10.1007/978-1-4614-1800-9_6}{Agent Based
  Modeling, Mathematical Formalism for}, pages 88--104.
\newblock Springer New York, 2012.

\bibitem{MortveitElements}
C.L. Barrett, H.S. Mortveit, and C.M. Reidys.
\newblock \href{http://dx.doi.org/10.1016/S0096-3003(00)00042-4}{Elements of a
  theory of simulation {III:} equivalence of {SDS}}.
\newblock {\em Applied Mathematics and Computation}, 122(3):325--340, 2001.

\bibitem{Aracena2004b}
J.~Aracena, J.~Demongeot, and E.~Goles.
\newblock {Fixed points and maximal independent sets in AND-OR networks}.
\newblock {\em Discrete Applied Mathematics}, 138:277--288, 2004.

\bibitem{Goles1982a}
F.~Fogelman-Soulie, E.~Goles-Chacc, and G.~Weisbuch.
\newblock {Specific roles of different Boolean mappings in random networks}.
\newblock {\em Bulletin of Mathematical Biology}, 44:715--730, 1982.

\bibitem{TCS2013}
M.~Noual, D.~Regnault, and S.~Sen{\'e}.
\newblock
  \mineb\href{http://www.sciencedirect.com/science/article/pii/S03043975120052%
82}{{A}bout non-monotony in Boolean automata networks}\minee.
\newblock {\em
  \href{http://www.journals.elsevier.com/theoretical-computer-science/}{Theore%
tical Computer Science}}, 504:12 -- 25, 2013.

\bibitem{AUTOMATA2010}
E.~Goles and M.~Noual.
\newblock
  \mineb\href{http://www.dmtcs.org/dmtcs-ojs/index.php/proceedings/article/vie%
wArticle/dmAL0104}{Block-sequential update schedules and Boolean automata
  circuits}\minee.
\newblock In {\em \href{http://automata.loria.fr/}{AUTOMATA}}, pages 41--50.
  \href{https://www.dmtcs.org/dmtcs-ojs/index.php/dmtcs/}{Discrete Mathematics
  \& Theoretical Computer Science (DMTCS)}, 2010.

\bibitem{MortveitSDS}
Henning~S. Mortveit and Christian~M. Reidys.
\newblock {\em An Introduction to Sequential Dynamical Systems}.
\newblock Springer-Verlag New York, Inc., 2007.

\bibitem{Perrot2015}
A.~Alcolei, K.~Perrot, and S.~Sen{\'e}.
\newblock \href{http://arxiv.org/abs/1510.05452}{On the flora of asynchronous
  locally non-monotonic Boolean automata networks}.
\newblock In {\em {Proceedings of SASB}}. Springer, 2015.
\newblock to appear.

\bibitem{AUTOMATA2012}
M.~Noual, D.~Regnault, and S.~Sen{\'e}.
\newblock
  \mineb\href{http://rvg.web.cse.unsw.edu.au/eptcs/paper.cgi?AUTOJAC2012.4}{Bo%
olean networks synchronism sensitivity and XOR circulant networks convergence
  time}\minee.
\newblock In {\em AUTOMATA \& JAC}, volume~90 of {\em
  \href{http://about.eptcs.org/}{Electronic Proceedings in Theoretical Computer
  Science (EPTCS)}}, pages 37--52. Open Publishing Association, 2012.

\bibitem{BenAmor2013}
H.~Ben~Amor, F.~Corblin, E.~Fanchon, A.~Elena, L.~Trilling, J.~Demongeot, and
  N.~Glade.
\newblock \href{http://dx.doi.org/10.1007/s10441-013-9169-5}{Formal Methods for
  Hopfield-Like Networks}.
\newblock {\em Acta Biotheoretica}, 61(1):21--39, 2013.

\bibitem{uncertainty2}
I.~Shmulevich, E.R. Dougherty, S.~Kim, and W.~Zhang.
\newblock Probabilistic boolean networks: a rule-based uncertainty model for
  gene regulatory networks.
\newblock {\em Bioinformatics}, 18(2):261--274, 2002.

\bibitem{uncertainty3}
R.~Dehghannasiri, B.{-}J. Yoon, and E.R. Dougherty.
\newblock \href{http://dx.doi.org/10.1186/s12859-015-0839-y}{Efficient
  experimental design for uncertainty reduction in gene regulatory networks}.
\newblock {\em {BMC} Bioinformatics}, 16:410, 2015.

\bibitem{Streck2012}
J.~Barnat, L.~Brim, A.~Krejci, A.~Streck, D.~Safranek, M.~Vejnar, and
  T.~Vejpustek.
\newblock \href{http://dx.doi.org/10.1109/TCBB.2011.110}{On Parameter Synthesis
  by Parallel Model Checking}.
\newblock {\em IEEE/ACM Trans. Comput. Biol. Bioinfo.}, 9(3):693--705, 2012.

\bibitem{Bernot2004}
G.~Bernot, J.-P. Comet, A.~Richard, and J.~Guespin.
\newblock Application of formal methods to biological regulatory networks:
  extending thomas' asynchronous logical approach with temporal logic.
\newblock {\em J. Theor Biol}, 229:339--347, 2004.

\bibitem{Vijesh2013}
N.~Vijesh, S.~Chakrabarti, and J.~Sreekumar.
\newblock Modeling of gene regulatory networks: A review.
\newblock {\em Journal of Biomedical Science and Engineering}, 6:223--231,
  2013.

\bibitem{ConfuseParallelAndSynchronism2}
F.~Hinkelmann, M.~Br, B.~Guang, R.~Mcneill, G.~Blekherman, A.~Veliz-cuba, and
  R.~Laubenbacher.
\newblock
  \href{http://bmcbioinformatics.biomedcentral.com/articles/10.1186/1471-2105-%
12-295}{ADAM: Analysis of Discrete Models of Biological Systems Using Computer
  Algebra}.
\newblock BioMed Central, 2011.

\bibitem{ConfuseParallelAndSynchronism3}
A.~Garg, A.~Di~Cara, I.~Xenarios, L.~Mendoza, and G.~De~Micheli.
\newblock \href{http://dx.doi.org/10.1093/bioinformatics/btn336}{Synchronous
  Versus Asynchronous Modeling of Gene Regulatory Networks}.
\newblock {\em Bioinformatics}, 24(17):1917--1925, 2008.

\bibitem{ConfuseParallelAndSynchronism4}
H.~De Jong{,}~D. Thieffry.
\newblock Mod{\'e}lisation, analyse et simulation des réseaux
  g{\'e}n{\'e}tiques.
\newblock {\em Med Sci}, 18, 2002.

\bibitem{ConfuseParallelAndSynchronism7Nice}
F.~Képès P.~Amar and V.~Norris, editors.
\newblock {\em Probabilistic Gene Network}. EDP Science, 2015.

\bibitem{ConfuseParallelAndSynchronism5Naldi}
A.~Naldi, P.T. Monteiro, and C.~Müssel {\it et al.}
\newblock Cooperative development of logical modelling standards and tools with
  colomoto.
\newblock {\em Bioinformatics}, 31:1154--1159, 2015.

\bibitem{Delaplace2010}
F.~Delaplace{,} H. Klaudel{,}~A. Cartier-Michaud.
\newblock Discrete causal model view of biological networks.
\newblock In {\em CMSB}, pages 4--13. ACM Press, 2010.

\bibitem{HDeJongReview}
H.~de~Jong.
\newblock \href{http://opac.inria.fr/record=b1043901}{Modeling and simulation
  of genetic regulatory systems : a literature review}.
\newblock Rapport de recherche INRIA, 2000.

\bibitem{ConfuseParallelAndSynchronism10}
C.~Chaouiya {\it et al}.
\newblock Sbml qualitative models: A model representation format and
  infrastructure to foster interactions between qualitative modelling
  formalisms and tools.
\newblock {\em BMC Systems Biology}, 7, 2013.

\bibitem{ConfuseParallelAndSynchronism11}
A.~Streck, T.~Lorenz, and H.~Siebert.
\newblock
  \href{http://dblp.uni-trier.de/db/journals/nc/nc14.html#StreckLS15}{Minimiza%
tion and equivalence in multi-valued logical models of regulatory networks}.
\newblock {\em Natural Computing}, 14(4):555--566, 2015.

\bibitem{ConfuseParallelAndSynchronism13Naldi}
C.~Chaouiya, A.~Naldi, and D.~Thieffry.
\newblock Logical modelling of gene regulatory networks with {GIN}sim.
\newblock {\em Methods in molecular biology}, 804:463--79, 2012.

\bibitem{ConfuseParallelAndSynchronism12}
C.J. Kuhlman{,} H.S. Mortveit{,} D. Murrugarra{,}~V.S.A. Kumar.
\newblock
  \href{http://www.dmtcs.org/dmtcs-ojs/index.php/proceedings/article/view/dmAP%
0103}{Bifurcations in Boolean Networks}.
\newblock In {\em AUTOMATA}, pages 29--46, 2011.

\bibitem{Thomas1990B}
R.~Thomas and R.~d'Ari.
\newblock {\em {Biological feedback}}.
\newblock CRC Press, 1990.

\bibitem{McCulloch1943}
W.~S. McCulloch and W.~H. Pitts.
\newblock {A logical calculus of the ideas immanent in nervous activity}.
\newblock {\em Bull Math Biophys}, 5:115--133, 1943.

\bibitem{Hopfield1982}
J.J. Hopfield.
\newblock Neural networks and physical systems with emergent collective
  computational abilities.
\newblock {\em ONAS USA}, 79(8):2554--2558, 1982.

\bibitem{IsingNeural}
H.G. Schaap.
\newblock {\em Ising models and neural networks}.
\newblock PhD thesis, Groningen, 2005.

\bibitem{Kauffman1969}
S.~A. Kauffman.
\newblock {Metabolic stability and epigenesis in randomly constructed genetic
  nets}.
\newblock {\em Journal of Theor. Biol.}, 22:437--467, 1969.

\bibitem{Thomas1973}
R.~Thomas.
\newblock {Boolean formalization of genetic control circuits}.
\newblock {\em J. Theor. Biol.}, 42:563--585, 1973.

\bibitem{Thomas1991}
R.~Thomas.
\newblock {Regulatory networks seen as asynchronous automata: a logical
  description}.
\newblock {\em J. Theoret. Biol.}, 153:1--23, 1991.

\bibitem{Caus}
M.~Noual.
\newblock \mineb\href{}{Causality and Boolean Automata Networks}\minee.
\newblock Research report, 2016.

\end{thebibliography}

\end{document}